\documentclass[12pt]{JHEP3}
\input epsf.tex
\epsfclipon

\usepackage{epsfig}
\usepackage{epstopdf}
\usepackage{epsf}
\input{epsf.sty}
\usepackage{graphicx,amsmath,amssymb}
\usepackage{epsfig,multicol}
\allowdisplaybreaks

\usepackage{bbm,bm,amsmath,amssymb}
\def\blfootnote{\xdef\@thefnmark{}\@footnotetext}

\long\def\symbolfootnote[#1]#2{\begingroup%
\def\thefootnote{\fnsymbol{footnote}}\footnote[#1]{#2}\endgroup}

\newcommand{\be}{\begin{eqnarray}}
\newcommand{\ee}{\end{eqnarray}}
\newcommand{\ben}{\begin{eqnarray*}}
\newcommand{\een}{\end{eqnarray*}}

\newcommand{\bcent}{\begin{center}}
\newcommand{\ecent}{\end{center}}
\newcommand{\benum}{\begin{enumerate}}
\newcommand{\eenum}{\end{enumerate}}
\newcommand{\bdesc}{\begin{description}}
\newcommand{\edesc}{\end{description}}
\newcommand{\bitem}{\begin{itemize}}
\newcommand{\eitem}{\end{itemize}}
\newcommand{\bquote}{\begin{quote}}
\newcommand{\equote}{\end{quote}}
\newcommand{\bhalfp}{\begin{minipage}{0.45\textwidth}}
\newcommand{\ehalfp}{\end{minipage}}
\newcommand{\bhead}{\begin{center}\bf \Large}
\newcommand{\ehead}{\end{center}\bigskip}

 \newcommand{\bfC}{{\bf C}}

 \newcommand{\bfP}{{\bf P}}

 \newcommand{\calO}{{\cal O}}

\newcommand{\p}{{\bf p}}

\def\be{\begin{equation}}
\def\ee{\end{equation}}
\def\ba{\begin{eqnarray}}
\def\ea{\end{eqnarray}}

\newcommand{\roughly}[1]{\mathrel{\raise.3ex\hbox{$#1$\kern-0.85em
\lower1ex\hbox{$\sim$}}}}

\def\2pi{\left(2\pi\right)}

\def\beq{\begin{equation}}
\def\eeq{\end{equation}}
\def\bg{\begin{eqnarray}}
\def\nd{\end{eqnarray}}
\def\bea{\begin{eqnarray}}
\def\eea{\end{eqnarray}}

\def\D3{\overline{\mbox{D3}}}

\def\nn{{\nonumber}}

\def\cO{{\cal O}}

\title{Supersymmetric Configurations, Geometric Transitions and New Non-K\"ahler Manifolds}

\author{Fang Chen$^1$, Keshav Dasgupta$^1$, Paul Franche$^1$, Sheldon Katz$^2$, Radu Tatar$^3$\\
\vskip.03in
${}^1$ Ernest Rutherford Physics Building, McGill University,\\
3600 University Street, Montr{\'e}al QC, Canada H3A 2T8\\
${}^2$ Departments of Mathematics and Physics,\\
University of Illinois at Urbana-Champaign,\\
1409 West Green Street, Illinois, USA 61801\\
${}^3$ Division of Theoretical Physics,\\
Department of Mathematical Sciences,\\
The University of Liverpool, Liverpool, England, UK L69 3BX\\
{\tt fangchen, keshav, franchep@hep.physics.mcgill.ca, katz@math.uiuc.edu, Radu.Tatar@liverpool.ac.uk}}
\date{November 2009}

\abstract{We give a detailed derivation of a supersymmetric configuration of wrapped D5-branes
on a two-cycle of a warped resolved conifold. Our analysis reveals that the resolved conifold should
support a non-K\"ahler metric with an $SU(3)$ structure. We use this as a starting point of the
geometric transition in type IIB theory.
A mirror, and a subsequent flop transition using an intermediate M-theory configuration with a $G_2$ structure,
gives rise to the complete IR geometric transition in type IIA theory.
A further mirror transformation gives the type IIB gravity
dual of the IR gauge theory on the wrapped D5-branes. Expectedly non-K\"ahler deformations of the resolved and the
deformed conifolds appear as the gravity duals of the confining gauge theories
in type IIA and type IIB theories respectively,
although in more generic cases these
manifolds could also be non-geometric. In the local limit we reproduce precisely the scenarios presented in
our earlier works. Our present work should therefore be viewed as providing a supergravity proof of geometric
transitions in the full global scenarios in type II theories.}

\maketitle

\begin{document}

\section{Introduction}

The gravity duals of gauge theories with running coupling constants have received
considerable attention in the last few
years. The original gauge/gravity duality \cite{Mal-1}, \cite{Witt-1} deals exclusively with theories that have no
running of the coupling constants, or with theories that have some running of the coupling constants but eventually
fall into fixed point surfaces, for example \cite{klebwitt}. The first kind of dualities that consider the actual
running of the coupling constants leading to, say, confining theories were discussed some time back in
\cite{KS}, \cite{vafa}, \cite{MN} and its extension to include fundamental flavors in \cite{ouyang}. The
type IIA brane constructions for theories like \cite{klebwitt} were first discussed in \cite{dasmukhi}, and for
theories with running couplings were discussed in \cite{ohtatar}. In fact in the fourth reference of \cite{ohtatar}
the precise distinctions between \cite{KS} and \cite{vafa} were pointed out in details.

In recent times we have seen many new advantages of studying theories like \cite{KS} and \cite{vafa} that deal with
running couplings. The confining behavior of these theories in the far IR is of course very powerful in extending them
to more realistic scenarios like high temperature QCD \cite{FEP, jpsi}. The cascading nature of these theories allow
them to remain strongly coupled throughout the RG flow from UV to IR, and therefore {\it supergravity} duals can
describe the full dynamics of the corresponding gauge theories. For the Klebanov-Strassler (KS) theory \cite{KS} even the
full UV completion, that allow no Landau poles or UV divergences of the Wilson loops, {\it can} be achieved by
attaching an UV cap to the KS geometry\cite{FEP}. An example 
of the full UV completion of the KS geometry both at
zero and non-zero temperatures has been recently accomplished in \cite{jpsi}. The UV cap therein is given by an
asymptotic AdS space that, in the dual gauge theory, will allow for an asymptotic conformal behavior in the UV and
linear confinement in the far IR.

For the model studied by Vafa \cite{vafa} the full UV completion would be more non-trivial. We expect the UV to be a
six-dimensional theory instead of a four-dimensional one. A six-dimensional UV completion that allows no Landau poles
in the presence of fundamental flavors has not been constructed so far.
In fact a proper study of fundamental flavors
{\it a-la} \cite{ouyang} is yet to be done for this case. The F-theory \cite{vafaF, DM, senF}
embedding of this model would be crucial
in analysing the full UV completion. However some aspects of an intemediate UV behavior, for example cascading dynamics,
have been discussed in the past \cite{katzvafa} where the cascade is likened to an infinite sequence of flop transitions.
The IR dynamics of the theory where we expect geometric transition to happen is actually the last stage
in this sequence of transformations
where the flop is immediately followed by a conifold transition. At this point we should
expect the wrapped D5-branes to be completely
replaced by fluxes (at least in the absence of fundamental flavors) \cite{katzvafa}.
What happens in the presence of fundamental flavors is rather subtle, and we will not discuss this here anymore. In fact
we will only concentrate on the last stage of the transition, namely, the geometric transition in this paper. The
intermediate cascading dynamics or the UV completion will be discussed elsewhere \cite{toappear}.

Since the geometric transition leads to a confining theory, the corresponding gauge dynamics is strongly coupled.
Therefore the physics of this transition can be captured exclusively by supergravity backgrounds. In some
of our earlier works \cite{gtpapers} we managed to study this purely using the supergravity backgrounds in the {\it local}
limit, meaning that the sugra background was studied around a specific chosen point in the internal six-dimensional space.
The reason for this was the absence of a known globally defined
supergravity solution of the wrapped D5-branes on the two-cycle of the
resolved conifold. The only known global solution i.e \cite{pandoz} was unfortunately not supersymmetric (see
\cite{cvetic, anke} for details) although it satisfied the type IIB EOMs. In this paper, among other things, we will
be able to solve this problem and provide a fully supersymmetric globally defined solution for the wrapped D5-branes on a
certain resolved conifold. What we will argue soon is that the resolved conifold should have a non-K\"ahler metric to
allow for supersymmetric solutions. This non-K\"ahlerity appears exactly from the back-reactions of the wrapped D5-branes.

Despite the absence of supersymetric solutions, in \cite{gtpapers} we managed to show, at least locally, the
full geometric transitons in type II theories. The gravity duals for the IR confining gauge theories on
the wrapped D6-branes in type IIA and wrapped
D5-branes in type IIB were completely captured by non-K\"ahler deformations of the resolved and the deformed conifolds
respectively. In this paper we will show that globally under some simplifying assumptions
this conclusion remains unchanged, but generically
these manifolds would become non-geometric (see \cite{halmagyi} for a recent discussion on this). In the following
sub-section we will briefly review the state of geometric transition using local supergravity analysis before we proceed
to compute the full global picture.

\subsection{Geometric transition and supersymmetric solution}

Let us begin with a bit of historical notes.
The original study of open-closed string duality in type II theory
starts with D6 branes wrapping a three cycle of a non-compact
deformed conifold. Naively one might expect the deformed conifold
to be a complex K\"ahler manifold with a non-zero three cycle.
However as discussed earlier in \cite{gtpapers} this is not quite correct,
and the manifold that actually would solve the string equations of
motion is a non-K\"ahler deformation of the deformed conifold. It
also turns out that the manifold has no integrable complex
structure, but only has an almost complex structure. This is
consistent with the prediction of \cite{vafaF}.

However, as one may recall, in all our earlier papers we managed to study only the
{\it local} behavior of the manifolds. This is because the full global picture was
hard to construct, and any naive procedure always tend to lead to non-supersymmetric
solutions. In deriving the local metric, we took a simpler model where all
the spheres were replaced by tori with periodic coordinates ($x,
\theta_1$) and ($y,\theta_2$). The coordinate $z$ formed a
non-trivial $U(1)$ fibration over the $T^2$ base. The replacement
of spheres by two tori was directly motivated from the
corresponding brane constructions of \cite{dasmukhi}, where
non-compact NS5 branes required the existence of tori instead of
spheres in the T-dual picture.

Locally
the non-K\"ahlerity of
the underlying metric can be easily seen from its explicit
form:
\bg\label{iiamet}
&&ds_{IIA}^2 = g_1~\left[(dz -
{b}_{z\mu}~dx^\mu) + \Delta_1~{\rm cot}~\hat\theta_1~(dx -
b_{x\theta_i}~d\theta_i) + \Delta_2~{\rm cot}~ \hat\theta_2~(dy -
b_{y\theta_j}~d\theta_j)+ ..\right]^2\nonumber\\
&& + g_2~ {[} d\theta_1^2
+ (dx - b_{x\theta_i}~d\theta_i)^2] + g_3~[ d\theta_2^2 + (dy -
b_{y\theta_j}~d\theta_j)^2{]} +  g_4~{\rm sin}~\psi~{[}(dx -
b_{x\theta_i}~d\theta_i)~d \theta_2 \nonumber\\
&& ~~~~~~~~~ + (dy -
b_{y\theta_j}~d\theta_j)~d\theta_1 {]} + ~g_4~{\rm
cos}~\psi~{[}d\theta_1 ~d\theta_2 - (dx - b_{x\theta_i}~d\theta_i)
(dy - b_{y\theta_j}~d\theta_j)]
\nd
where the coefficients $g_i$
and the coordinates $\theta_i, \hat\theta_i$ etc. are defined in
\cite{gtpapers}. The background has
non-trivial gauge fields (that form the sources of the wrapped D6
branes) and a non-zero string coupling (which could in principle
be small).

Existence of such an exact supergravity background helps us to obtain
the corresponding mirror type IIB background. One would expect that
this can be easily achieved
using the mirror rules of \cite{syz}. It turns out however that
the mirror rules of \cite{syz}, as discussed in \cite{gtpapers}, do not
quite suffice.
A detailed analysis of this is
presented in \cite{gtpapers}.
As discussed therein, we have to be careful
about various subtle issues while doing the mirror transformations:

\noindent (a) The mirror rules of \cite{syz} tells us that {\it any} Calabi-Yau
manifold with a mirror admits, at least {\it locally}, a $T^3$ fibration over a
three dimensional base. This seems to fail for the deformed
conifold as it does not possess enough isometries to represent it as a $T^3$
fibration.
On the other hand, a resolved conifold does have a well defined $T^3$
torus over a three-dimensional base, which can be exploited to get the mirror (see also
\cite{agavafa}). It also turns out that the $T^3$ torus is a lagrangian
submanifold, so a mirror transformations will not break any
supersymmetry.

\noindent (b) Viewing the mirror transformation naively as three T-dualities
along the $T^3$ torus {\it does~not} give the right mirror metric. There are
various issues here. The rules of \cite{syz} tell us that the mirror transformation
would only work when the three dimensional base is very large. The configuration that
we have is exactly opposite of the case \cite{syz}. Recall that
our configuration lies at the end of a much larger cascading theory.
By UV/IR correspondences, this means that the
base manifold is very small. Furthermore we are at the {\it tip} of the
geometric transition and therefore we have to be in a situation with very
small base (in fact very small fiber too). In \cite{gtpapers} we showed
that we could still apply the rules of \cite{syz} if we impose a non-trivial large
complex structure on the underlying $T^3$ torus. The complex structure
can be integrable or non-integrable. Using an integrable complex structure, we
showed in \cite{gtpapers} that we can come remarkably close to getting
the right mirror metric. Our conjecture there was that if we use a
non-integrable complex structure we can get the right mirror manifold.

It seems therefore natural to start with the manifold that exhibits three isometry
directions --- the resolved conifold. We can, however, not use the metric for D5
branes wrapping the $S^2$ of a resolved conifold as derived in \cite{pandoz}, because
it breaks all supersymmetry \cite{cvetic}.  The metric that we
proposed in \cite{gtpapers} (where we kept the harmonic functions undetermined) is very
close to the metric of \cite{pandoz} but differs in some subtle way:

\noindent (a) The type IIB resolved conifold metric that we proposed in
\cite{gtpapers} is a D5 wrapping a two cycle that {\it preserves} supersymmetry.
We will discuss this issue in more detail below.

\noindent (b) As explained in \cite{gtpapers}, our IIB manifold
also has seven branes (and possibly orientifold planes) along with
the type IIB three-form fluxes. The metric constructed in \cite{pandoz}
doesn't have seven branes but allows three-form fluxes.

\noindent The {\it local} behavior of the type IIB metric is expressed in terms
of non-trivial complex structures $\tau_1$ and $\tau_2$ as $dz_1 = dx - \tau_1 d\theta_1$ and
$dz_2 = dy - \tau_2 d\theta_2$. The local metric then reads
\bg\label{iibmet}
ds^2 = (dz + \Delta_1 ~ {\rm cot}~\theta_1 ~ dx  +
\Delta_2 ~{\rm cot}~\theta_2 ~dy)^2 + \vert dz_1 \vert^2 + \vert dz_2\vert^2
\nd
where all the warp factors can locally be absorbed in to the coordinate differentials.
In this formalism the metric may naively look similar to the one studied in
\cite{pandoz} but the global picture is completely different from the
one proposed by \cite{pandoz}. Our aim in this paper is therefore two-fold: to determine the full global picture (at
least without the inclusion of UV caps), and to follow the duality cycle that will lead us to analyse
geometric transitions in type II theories.

\subsection{Organisation of the paper}

The paper is organised as follows. In section 2
we start with geometric preliminaries about the resolved conifold and the
blown-up conifold.  We then discuss the mathematical construction of a non-K\"ahler resolved
conifold with an $SU(3)$ structure, following \cite{llt}.
%In this section we also analyse the torsional equations that will allow for
%such a solution to exist. Solutions of these torsional equations are presented
The components of the metric of this $SU(3)$ structure
(which automatically satisfy the
torsional equations) are given in {\bf Appendix 1}, and we use
these components to determine the metric of the wrapped D5-branes on the non-K\"ahler resolved conifold in
section 4.2. Existence of an $SU(3)$ structure will guarantee that the solution we get is supersymmetric, and we
discuss the issue of supersymmetry further in section 4.1.

Sections 4.3 to 4.6 are the main sections where we compute the full geometric transitions in the global framework
using duality cycle that were used earlier in \cite{gtpapers}. In the full global picture the fluxes are very involved
compared to the local picture. We managed to work out all the fluxes in the type IIA mirror set-up. These flux
components are given in {\bf Appendix 2}. It is necessary to track these fluxes because they would eventually determine
the non-K\"ahler fibration structure in type IIB theory discussed in section 4.6. We discuss the components of the
metric in type IIB after geometric transition in {\bf Appendix 3}.

In the above discussion we have briefly alluded to the fundamental flavors. In our model they appear from orientifolding
the resolved conifold. The orientifolding is subtle, and we give a brief discussion of this in section 3. This
orientifolding will allow us to add seven-branes in type IIB and six-branes in the mirror type IIA picture. The
seven-branes should be embedded as in \cite{pandoz} or \cite{kuperstein}.

We end with a conclusion and a short discussion on the topics that we will study in \cite{toappear}.

\section{Mathematical constructions and $SU(3)$ structures}

The blown-up conifold describes one of the topologies that we will be using.  We will describe it as well as its relation to the resolved conifold which is more familiar
in string theory.  These ideas are
well known in algebraic geometry.  Rather than starting with the standard algebro-geometric constructions, we instead begin with a description more suited
to the description of an $SU(3)$ structure.  The algebro-geometric description
will follow.

\subsection{The blown-up conifold}
\label{blowup}

Let us begin by explaining how the blown-up conifold arises for us.  The
conifold is a cone over $S^3\times S^2$ \cite{candelas}, arising as a quotient
$(S^3\times S^3)/U(1)$, with the $U(1)$ diagonally embedded and on each
factor identified with the $U(1)$ of the Hopf fibration $S^3\to S^2$.
If we take the self-product of the Hopf fibration
$S^3\times S^3\to S^2\times S^2$ and mod out by the diagonal $U(1)$, we are
left with a $U(1)$ fibration $S^3\times S^2\to S^2\times S^2$ which restricts
to the Hopf fibration over either $S^2$ factor.

In the geometry we will be using which supports the metric (\ref{susybg}),
the internal space is a 6-manifold, the total space of a complex line bundle
$L$ with base $S^2\times S^2$ which is described in spherical coordinates
$(\theta_i,\phi_i)$ for $i=1,2$.  The fiber is described by an angular coordinate $\psi$ describing the nontrivial $U(1)$ bundle
over $S^2\times S^2$ just described.  The $U(1)$ bundle is completed to
a complex line bundle by introducing the radial coordinate $r$.

The topology of either the line bundle or the $U(1)$ bundle is described completely by its Chern class on $S^2\times S^2$.
As noted above, the
$U(1)$ bundle restricted to either $S^2$ is the Hopf bundle $S^3$, whose Chern class on $S^2$ is well known to have degree $-1$.  So on
$S^2\times S^2$, we learn that $L$ has degrees $(-1,-1)$.

We can now identify $S^2$ with the complex projective line $\bfP^1$ and switch to the language of algebraic geometry, whereby we see that the internal manifold $X$ is the total space of the line
bundle $\calO(-1,-1)$ on $\bfP^1\times\bfP^1$.

This manifold appears as
the blown-up conifold in algebraic geometry.  Rather than refer to known results, we prefer to directly identify $X$ with the blown-up conifold by describing the map from $X$ to the conifold which shrinks $\bfP^1\times\bfP^1$ (identified with the zero section of $L$) to the conifold point.

We introduce homogeneous coordinates $(u^1,u^2)$ and $(v^1,v^2)$ on the respective
$\bfP^1$'s, and identify sections of $L$ with functions on $\bfP^1\times\bfP^1$ which are homogeneous of degree $-1$ with respect to $(u^1,u^2)$ as well as with respect to $(v^1,v^2)$.  Thus a point of $X$ can be described by homogeneous coordinates
$(u^1,u^2,v^1,v^2,s)$, where $s$ is thought of as a section of $L$.  The homogeneity is
described by two $\bfC^*$ actions whose respective weights are
$(1,1,0,0,-1)$ and $(0,0,1,1,-1)$.

The map from $X$ to the conifold is realized by the map from $X$ to $\bfC^4$ given by

\bg\label{bdtoconifold}
(u^1,u^2,v^1,v^2,s)\mapsto (u^1v^1s,u^2v^2s,u^1v^2s,u^2v^1s).
\nd

If we introduce coordinates $(x_1,\ldots,x_4)$ on $\bfC^4$, we see that the image of
$X$ satisfies the equation
\bg
x_1x_2-x_3x_4=0
\nd
of the conifold.  Identifying $\bfP^1\times\bfP^1$ with the zero section $s=0$, we see
that $\bfP^1\times\bfP^1$ is collapsed to the conifold point $(0,0,0,0)$ as claimed.

When described by homogeneous coordinates as above, Calabi-Yau manifolds are characterized by the condition that the sum of the weights is zero for any $\bfC^*$. Since the sum of the weights is one for either $\bfC^*$, we conclude that the blown-up conifold is not a Calabi-Yau manifold.  We will also check this
directly in the next section by the adjunction formula.

\subsection{The resolved conifold}\label{rescon}

We can relate the blown-up conifold to the more familiar resolved conifold.
Rather than blow down $\bfP^1\times\bfP^1$ to the conifold point as in (\ref{bdtoconifold}), we can instead partially blow down $\bfP^1\times\bfP^1$ by
projecting to one $\bfP^1$. This gives the usual resolved
conifold.

\smallskip
Using the coordinates of the blown-up conifold introduced in Section~\ref{blowup}, the partial blowdown is described by

\bg\label{bdtoresolved}
(u^1,u^2,v^1,v^2,s)\mapsto
(u^1,u^2,v^1s,v^2s),
\nd
so that $\bfP^1\times\bfP^1$, identified with $s=0$ as before, is mapped to
$(u^1,u^2,0,0,0)$, and $\bfP^1\times\bfP^1$ is projected to the first coordinate, as
claimed. We let $(z^1,z^2,z^3,z^4)$ be homogeneous coordinates on the image of (\ref{bdtoresolved}).

Only one $\bfC^*$ remains nontrivial on $(z^1,z^2,z^3,z^4)$, with weights $(1,1,-1,-1)$.  The image is the resolved conifold, the total space of $\calO(-1)\oplus\calO(-1)$ on $\bfP^1$.  The coordinates $(z^1,z^2)$ can be identified
with the homogeneous coordinates of $\bfP^1$, while $z^3$ and $z^4$ are identified with
sections on the respective copies of $\calO(-1)$.   The resolved conifold is of course
Calabi-Yau, which can be seen since the sum of the weights is 0.

\bigskip
We remark that we have taken a circuitous path to get from
the blown-up conifold to the more familiar resolved conifold, but we
have reached the usual descriptions of the resolved conifold as either
the total space of the bundle $\calO(-1)\oplus\calO(-1)$ on $\bfP^1$
or as a toric quotient of $\bfC^4$ by $\bfC^*$ with weights $(1,1,-1,-1)$.

\bigskip
However, it is not the Calabi-Yau structure that is relevant in our model, but rather
a non-K\"ahler structure.  To construct this non-K\"ahler structure, the holomorphic
homogeneous coordinates are not particularly useful.
At times it will be useful to describe the resolved conifold as a
symplectic quotient, or equivalently as the space of vacua of a $U(1)$ gauge theory with four scalar fields with $U(1)$ charges $(1,1,-1,-1)$, whose
vevs are identified with $(z^1,z^2,z^3,z^4)$.  There is an FI term $u$, which we take to be positive.  If we take $u<0$, we get the flopped version of the resolved conifold.

So the resolved conifold can be described by
\bg\label{symprescon}
|z^1|^2+|z^2|^2-|z^3|^2-|z^4|^2=u
\nd
modulo the $U(1)$ action.

We can see directly from this description that the resolved conifold is smooth.  We can for
example describe the patch in which $z^1\ne0$ by the complex coordinates $(z^2,z^3,z^4)$
and solve (\ref{symprescon}) by
\bg\label{resconcoords}
z^1=\sqrt{u-|z^2|^2+|z^3|^2+|z^4|^2}.
\nd
Note that in (\ref{resconcoords}) we have fixed the gauge by choosing the positive real
solution of (\ref{symprescon}), so that $(z^2,z^3,z^4)$ are in fact local coordinates.
However, they are in no sense to be considered as holomorphic coordinates since
(\ref{resconcoords}) is not holomorphic.

\bigskip
We now turn to the holomorphic description.  Start with
the conifold singularity $X$ with
equation
$$x_1x_2-x_3x_4=0$$
which we denote by $f=0$.
The usual resolved conifold $X'$ can be described as the submanifold of
$\bfC^4\times \bfP^1$ with equations which we informally write as
$$\frac{x_1}{x_3}=\frac{x_4}{x_2}=\frac{y_2}{y_1},$$ or more formally as
$$x_1x_2-x_3x_4=0,\ x_1y_1=x_3y_2,\ x_4y_1=x_2y_2.$$  In the above,
$(y_1,y_2)$ are the homogeneous coordinates on $\bfP^1$.

If $x=(x_1,x_2,x_3,x_4)\ne (0,0,0,0),$ then there is a unique solution for
$y=(y_1,y_2)$, so that $X$ and $X'$ are isomorphic away from the origin.  If
however, $x=0$, then $y$ is unconstrained and we replace the origin by
a $\bfP^1$ to form $X'$ from $X$.

This $\bfP^1$ can be flopped to produce another resolved conifold $X''$.
The flop can be realized directly by the equations
$$\frac{x_1}{x_4}=\frac{x_3}{x_2}=\frac{y_2}{y_1}.$$
$X'$ and $X''$ are isomorphic in this local model, but in global models
$X'$ and $X''$ containing a resolved conifold and its flop respectively, the
Calabi-Yaus $X'$ and $X''$ need not be isomorphic.

\bigskip
In this paper, we use the model where both $\bfP^1$'s are introduced
simultaneously.  This is accomplished by the algebro-geometric construction
of blowing up the conifold, which complements our description in Section 2.1. We introduce a $\bfP^3$ with homogeneous
coordinates $(y_1,y_2,y_3,y_4)$ and the blowup $\tilde{X}$ is constructed
as the submanifold of $\bfC^4\times\bfP^3$ with equation informally expressed as
$$(x_1,x_2,x_3,x_4)=(y_1,y_2,y_3,y_4).$$  As before, if $x\ne0$ then there is
a unique solution for $y$ and so $\tilde{X}$ is isomorphic to $X$ away from
the origin.  If $x=0$, there are more solutions, but now there is a
constraint $y_1y_2=y_3y_4$.  This is a quadric surface in $\bfP^3$, isomorphic
to $\bfP^1\times \bfP^1$ by the isomorphism
$$\left((u_1,u_2),(v_1,v_2)\right)\mapsto (u_1v_1,u_2v_2,u_1v_2,u_2v_1).$$

Note that the blown-up conifold
$\tilde{X}$ is not Calabi-Yau, but we can realize
this within string theory by turning on an appropriate flux.

\smallskip
We consider the holomorphic 3-form $\Omega$ on $X$ given by the usual
residue construction
$$\Omega=\frac{dx_2\wedge dx_3\wedge dx_4}
{\partial f/\partial x_1}=\frac{dx_2\wedge dx_3\wedge dx_4}{x_2}$$ and pull it back to a holomorphic
3-form $\tilde{\Omega}$ on $\tilde{X}$.
To see that $\tilde{X}$ is not Calabi-Yau, it suffices to show that
$\tilde\Omega$ vanishes somewhere on $\tilde{X}$. It
suffices to compute in one coordinate patch, say where $x_1\ne0$.  In
this patch, $(x_1,y_3,y_4)$ are local coordinates.  To see this, we may
set $y_1=1$, and then
we use $y_i=x_i/x_1$ for $i=2,3,4$ to compute
$$x_2=x_1y_3y_4,\ x_3=x_1y_3,\ x_4=x_1y_4.$$
In these coordinates, we have $\tilde{\Omega}=x_1dx_1\wedge dy_3\wedge dy_4$,
which clearly vanishes on the surface $x_1=0$.
Thus $\tilde{X}$ is not Calabi-Yau.

For later use, note that in this coordinate patch, if $x_1=0$ then necessarily
$x_2=x_3=x_4=0$ as well.  Thus $x_1=0$ is the local equation of the
exceptional $\bfP^1\times\bfP^1$  of $\tilde{X}$, which we denote by $E$.

Alternatively we can see the same result by the adjunction formula \cite{gh}, which
says that for any hypersurface $H$ in a complex manifold $M$, we have
$K_H=(K_M+[H])\mid_H$.

We realize $\tilde{X}$ as a hypersurface in the blowup of $\bfC^4$ at a point
and apply the adjunction formula.

Let $E'$ be the $\bfP^3$ which is the exceptional divisor of the
blowup $Z$ of $\bfC^4$ at the origin; then $K_Z=3E'$ \cite{gh}.  Then the
adjunction formula yields
$$K_{\tilde{X}}=\left(K_Z+\tilde{X}\right)|_{\tilde{X}}.$$

Now $\tilde{X}$ is obtained by subtracting off the exceptional divisor from
the pullback of $X$ via the blowup.  Since $X$ has a multiplicity~2 singularity
at the origin, the pullback of $X$ actually contains the exceptional
divisor $E'$ with multiplicity~2.
Since $2E'$ has to be subtracted off to obtain $\tilde{X}$, we conclude
that $\tilde{X}$  has divisor class $-2E'$.  We
conclude that
$$K_{\tilde{X}}=\left(3E'-2E'\right)|_{\tilde{X}}=E'|_{\tilde{X}}=E,$$
so that $K_{\tilde{X}}$ is nontrivial and $\tilde{X}$ is not Calabi-Yau.

This is consistent with the explicit calculation.  Since $x_1=0$ defines
the exceptional divisor $E$, the fact that $\tilde{\Omega}$ vanished precisely
on $E$ tells us that $K_{\tilde{X}}=E$.

\subsection{$A_1$ fibered geometry}

We start with the blown up $A_1$ geometry fibered over the complex numbers
with parameter $x_4\in\bfC$.  The equation is just the $A_1$ equation
\bg\label{a1}
x_1x_2-x_3^2=0
\nd
and the blowup is performed by introducing a $\bfP^1$ with homogeneous
coordinates $(y_1,y_2)$ and imposing the equations
$$\frac{x_3}{x_1}=\frac{x_2}{x_3}=\frac{y_2}{y_1},$$
or more formally
\bg\label{resolveda1}
x_3y_1=x_1y_2,\qquad x_2y_1=x_3y_2.
\nd
Note that $x_4$ does not appear explicitly, and can be interpreted as a
parameter for the
location of $\bfP^1$.  The $\bfP^1$ corresponding to $x_4=\phi$ will be written
as $C_\phi$.

For later use, the normal bundle of $C_\phi$ is $\cO_{C_\phi}\oplus
\cO_{C_\phi}(-2)$.

We now deform this geometry with deformation parameter $t$:
\bg\label{a1deformed}
x_1x_2-x_3^2+t^2x_4^{2n}=0
\nd
and the blowup is performed by introducing a $\bfP^1$ with homogeneous
coordinates $(y_1,y_2)$ and imposing the equations
$$\frac{x_3-tx_4^n}{x_1}=\frac{x_2}{x_3+tx_4^n}=\frac{y_2}{y_1}.$$

This geometry corresponds to the superpotential $W(\phi)=t\phi^{n+1}/(n+1)$.
Note that for $t=0$ we have $W(\phi)\equiv0$, and the curve $C_\phi$ is
holomorphic for all $\phi$.  For $t\ne0$, we have $W'(\phi)=t\phi^n$, and
$C_\phi$ only  persists holomorphically for $\phi=0$.  To see this, the
requirement is
$$x_3-t\phi^n=x_1=x_2=x_3+t\phi^n=0,$$
which implies that $\phi^n=0$.

The superpotential can be obtained by integrating the holomorphic
3-form $\Omega=dx_2dx_3dx_4/x_2$ over a three chain $\Gamma$
connecting $C_0$ to $C_\phi$.  This can be reinterpreted in terms of
the relative homology class of $\Gamma$.  The same relatively
cohomology group can be realized after blowing up $C_0$ and $C_\phi$.

Initially putting $t=0$, the blowup of $C_0$ has exceptional divisor isomorphic
to the Hirzebruch surface $F_2$.  A similar blowup can be performed on
$C_\phi$.

The Hirzebruch surface $F_2$ deforms if $\phi=0$ but not otherwise.  If
$n>1$, the deformed surface is still $F_2$.  If $n=1$, then the deformed
surface is $\bfP^1\times\bfP^1$.

If desired, a toric description of the blowup can be given.  The $A_1$
surface singularity is a toric variety whose fan has a single
two-dimensional cone with edges spanned by $(1,0)$ and $(1,2)$.  The
singularity gets resolved by inserting an extra edge $(1,1)=(1/2)
((1,0)+(1,2)$.  This resolved $A_1$ gets fibered over $\bfC$ in the
usual way: by adding another coordinate, appending a zero to the
coordinates of the vectors spanning the edges, and adding the new
coordinate vector.  Hence the fan has edges spanned by

$$(1,0,0), (1,1,0), (1,2,0), (0,0,1).$$

The curve $C_0$ corresponds to the 2-dimensional cone spanned by $(1,1,0)$
and $(0,0,1)$, so $C_0$ gets blown up by inserting a new edge spanned by
$(1,1,1)=(1,1,0)+(0,0,1)$.  In summary, the toric variety has edges

$$(1,0,0), (1,1,0), (1,2,0), (0,0,1), (1,1,1).$$

\subsection{$SU(3)$ structure}\label{su3}
We follow \cite{llt} which gives a general procedure for constructing string
compactifications on toric varieties via a method for producing
$SU(3)$ structures.  Any $SU(3)$ structure arises as a string compactification \cite{torsion}.

We apply the method to the resolved conifold.  The method
was designed to apply to compact toric varieties, but since the method has a
local character, it may be applied to the resolved conifold.  We set ourselves
to that task.

For the convenience of the reader, we collect some facts about $SU(3)$
structures.

\bigskip
An $SU(3)$ structure on a 6-manifold $M$
is determined by a complex decomposible 3-form $\Omega$ and a real 2-form $J$ which are
related by
\bg\label{su3conds}
\Omega\wedge J=0, \qquad \Omega\wedge\overline{\Omega}=-\frac{4i}3J\wedge J\wedge J.
\nd

At each point $p\in M$, we can find complex cotangent vectors $dz^1,dz^2,dz^3$ so that $\Omega=
dz^1\wedge dz^2\wedge dz^3$ at $p$.  The first condition of (\ref{su3conds}) and the
reality of $J$ imply that we can ``diagonalize" $J$, writing it as
$$J=\frac{i}2\left(a_1dz^1\wedge dz^{\bar{1}}+a_2dz^2\wedge dz^{\bar{2}}+a_3dz^3\wedge dz^{\bar{3}}\right)$$
for some real constants $a_i$, while retaining the form of $\Omega$.  Then the second condition of (\ref{su3conds})
implies that we can rescale the $dz^i$ so that
\bg
J=\frac{i}2\left(dz^1\wedge dz^{\bar{1}}+dz^2\wedge dz^{\bar{2}}+dz^3\wedge dz^{\bar{3}}\right)
\nd
while $\Omega=dz^1\wedge dz^2\wedge dz^3$ still holds.

There is still the freedom of
multiplying the $dz^i$ by phases whose product is 1.

These coordinates determine a Euclidean metric $g_{i\bar{j}}=i/2$ at $p$, which
is independent of the phase ambiguity.

This pointwise analysis extends to all of $M$, showing that the data of $\Omega$ and
$J$ satisfying (\ref{su3conds}) completely determines a metric, the metric associated
with an $SU(3)$ structure.

There is an explicit procedure to calculate the metric directly from $\Omega$ and $J$.
The first step is to calculate the (not necessarily integrable) complex structure $I$.  In pointwise Euclidean coordinates at $p$, the complex structure $I$ is the standard one.  But it can be computed intrinsically following 
\cite{hitchin} as follows\footnote{For more details on the following analysis, and also to connect to recent conifold 
literature the readers may refer to \cite{papat} and references therein.}.

First define an unnormalized complex structure by
\bg
\tilde{I}_j^k=
\epsilon^{klmnop}\left(\mathrm{Re}\Omega\right)_{jlm}\left(\mathrm{Re}\Omega\right)_{nop},
\nd
where $\epsilon$ is the completely antisymmetric tensor.  It is shown in \cite{hitchin}
that $\tilde{I}^2$ is a diagonal matrix with negative real entries.  Then
\bg
I=\frac{\tilde{I}}{\sqrt{-\frac16\mathrm{Tr}\tilde{I}^2}}
\nd
is the desired complex structure.

From here, the metric is determined by
\bg
g_{ij}=I^k_jJ_{ki}.
\nd
{}From the definitions, $\Omega$ has type $(3,0)$ and $J$ has type $(1,1)$
in the complex structure $I$.
If $(g,I)$ determines a K\"ahler structure, then $J$ is just the usual
K\"ahler form.

\bigskip
We also quickly review the method of \cite{llt} while applying it to the resolved
conifold.  A general method is developed for producing $SU(3)$ structures on
three-dimensional complex toric varieties, by producing an $\Omega$ and $J$ satisfying
(\ref{su3conds}).  The novel ingredient is to produce a $(1,0)$ form on complex
Euclidean space satisfying certain conditions, and then then $SU(3)$ structure is
determined by formulae.

We recall that the resolved conifold has been described as a quotient of $\bfC^4$ by
the $\bfC^*$ with weights $Q=(1,1,-1,-1)$.
So it suffices to produce a $(1,0)$ form
$K=K_idz^i$ on $\bfC^4$ satisfying (3.15), (3.16) and the normalization condition
(3.18) of \cite{llt}.  We interpret these conditions in concrete terms.

The condition (3.15) is equivalent to $Q^iz^iK_i=0$.  The condition (3.16)
says that $K$ has half the $U(1)$ charge as the holomorphic volume form
$\Omega_{\bfC}=dz^1\cdots dz^4$ of $\bfC^4$, which is zero in this case.  Our normalization condition is $\sum |K_i|^2 =1$, slightly different from
(3.18) but we will adjust for it later.

An obvious solution is

\bg\label{kdefn}
K=\frac{z^3dz^1+z^1dz^3+z^4dz^2+z^2dz^4}{|z|}.
\nd

We now compute the $SU(3)$ structure, following the formulas in \cite{llt}.  We first construct
the standard $SU(3)$ structure on the resolved conifold before modifying it.  The
$\bfC^*$ action on $\bfC^4$ is generated by
\bg
V=z^1\partial_{z^1}+z^2\partial_{z^2}-z^3\partial_{z^3}-z^4\partial_{z^4}
\nd
and then the standard Calabi-Yau 3-form is
\bg
\widetilde{\Omega}&=&~i_V\Omega_{\bf C}\\
&=&~z^1dz^2\wedge dz^3\wedge dz^4-z^2dz^1\wedge dz^3\wedge
dz^4-z^3dz^1\wedge dz^2\wedge dz^4+z^4dz^1\wedge dz^2\wedge dz^3.\nonumber
\nd
The K\"ahler form $\tilde{J}$ of the resolved conifold arises by modifying the K\"ahler form of
$\bfC^4$
\bg
J_{\bfC}=\frac{i}2\sum_{i=1}^4dz^i\wedge dz^{\bar{i}}
\nd
by putting
\bg
\eta=\bar{z}^1dz^1+\bar{z}^2dz^2-\bar{z}^3dz^3-\bar{z}^4dz^4,
\nd
with the coefficients coming from the weights, and then putting
\bg
\tilde{J}=\frac1{|z^2|}\left(J_{\bfC}-\frac{i}2\eta\wedge\bar{\eta}\right).
\nd
Note that $\eta$ respects the $U(1)$ action but not the $\bfC^*$ action.  So from this
point forward, we have to understand differential forms as $U(1)$-invariant forms,
subject to the D-term constraint (\ref{symprescon}).  In particular, the K\'ahler
form implicitly depends on the FI parameter $u$, as it must.

%$$V=z^1\frac{\partial}{\partial z^1}-z^2\frac{\partial}{\partial z^2}
%+z^3\frac{\partial}{\partial z^3}-z^4\frac{\partial}{\partial z^4}$$
%$$g=|z|^2,\qquad \tilde{g}=|z|^{-2}$$
%which gives
%$$
%\begin{array}{ccc}
%P(dz^1)&=&dz^1-\frac{z^1}{|z|^2}\eta\\
%P(dz^2)&=&dz^2+\frac{z^2}{|z|^2}\eta\\
%P(dz^3)&=&dz^3-\frac{z^3}{|z|^2}\eta\\
%P(dz^4)&=&dz^4+\frac{z^4}{|z|^2}\eta
%\end{array}
%$$
%Put
%$$K=\frac{z^2dz^1+z^1dz^2+z^3dz^4+z^4dz^3}{\left(2|z|^2\right)^{1/2}}$$
%Clearly (3.15) $P(K)=K$ is satisfied since the coefficients have been arranged
%so the $\eta$ terms drop out.  Since
%$\Omega_{\bfC} =dz^1\wedge dz^2\wedge dz^3\wedge dz^4$ has charge 0, (3.16) is
%the requirement that $K$ has charge zero, which is clearly fulfilled.
%Finally, the normalization has been chosen to ensure (3.18).

\bigskip
We now use $K$ to modify $\tilde{\Omega}$ and $\tilde{J}$, effectively replacing
$K$ by $\bar{K}$ throughout.

\smallskip
An auxiliary $SU(2)$ structure is created, characterized by two-forms $\omega$ and $j$ satisfying
\bg\label{su2cond}
\omega\wedge j=0,\qquad \omega\wedge\bar{\omega}=2j\wedge j,
\nd
where
\bg
j=\tilde{J}-\frac{i}2K\wedge\bar{K}
\nd
and $\omega$ is given by
\bg\label{su2}
\omega_{ij}=-2\bar{K_l}\eta^{m\bar{l}}\tilde{\Omega}_{mij},
\nd
where $\eta$ is the Euclidean metric on $\bfC^4$. The prefactor of 2 here on the right-hand side of (\ref{su2}) is not present in \cite{llt} but is required by our
normalization condition for $K$.

From here, we get a 2-parameter family of $SU(3)$ structures given by
\bg
J=aj-\frac{ib^2}2K\wedge\bar{K},\qquad
\Omega=ab\bar{K}\wedge\omega.
\nd
Using (\ref{su2cond}), it is immediate to see that $J$ and $\Omega$ satisfy the conditions (\ref{su3conds}) for an $SU(3)$ structure.

\smallskip
There is an extra phase parameter for $\Omega$ in \cite{llt}, but we supress it here
since the metric does not depend on this phase. In {\bf Appendix 1} we write down all the components of the metric.

\section{T-duality and Orientifold Projection}

The  blown-up conifold that we discussed above, has a product structure of
$\bfP^1\times\bfP^1$ and we would like to discuss the T-duality to a IIA model. Our final aim is to see how
orientifold projection effects the T-duality.
To do so we first review the case of \cite{orientifold} for the resolution of a deformed $A_2$ singularity.
This involves a
natural way to introduce two $\bfP^1$ cycles.

\subsection{Brief review of the deformed $A_2$ case}

Let us consider  the singular space $X_0$ realized as:
\be
\label{X0}
xy=(u-t_0(z))(u-t_1(z))(u-t_2(z))~~, \ee where $x,y,u,z$ are the
affine coordinates of $C^4$ and  $t_j(z)$ are polynomials.

The affine variety (\ref{X0}) has $A_1$ singularities at
$x=y=0$ and $z$ one of the double points of the planar
algebraic curve: \be
\label{Sigma_0}
\Sigma_0:~~(u-t_0(z))(u-t_1(z))(u-t_2(z))=0~~.  \ee
The curve has 3 components $C_j$ given by
$u=t_j(z)$, each a section of the $A_2$ fibration (\ref{X0}).

The resolved space ${\hat X}$ can be described explicitly as
follows. Consider two copies of
$\bfP^1$ with homogeneous coordinates $[u_1,u_2]$ and $[v_1,v_2]$, respectively, and local
affine coordinates $\xi_1:=u_1/u_2, \xi_2:=v_1/v_2$.  Then
${\hat X}$ is realized as :
 \bea
\label{hatX}
u_2 (u-t_0(z))&=&u_1 x\nn\\
v_1(u-t_1(z))&=&v_2 y
\eea

The IIA construction is obtained by  performing a T-duality with respect to the
following $U(1)$ action on ${\hat X}$, which we denote by
 \be
\label{rho_global}
([u_1,u_2], [v_1,v_2],z,u,x,y)\stackrel{{\hat \rho}(\theta)}{\longrightarrow}
([e^{-i\theta}u_1,u_2],
[v_1,e^{i\theta}v_2],z,u,e^{i\theta}x,e^{-i\theta} y)~~.  \ee
This projects as follows on the singular space $X_0$:
\be
\label{U1proj}
(z,~u,~x,~y)\stackrel{\rho_0(\theta)}{\longrightarrow} (z,~u,~e^{i\theta}x,~e^{-i\theta}y)~~.
\ee

In the type IIB set-up, we consider the case
\bea
\label{tspec}
t_0(z)&=&t(z)\nn\\
t_1(z)&=&t(-z)\\
t_2(z)&=&-t(z)-t(-z)\nn~~.  \eea

In this situation, the resolution ${\hat X}$ admits a $Z_2$ symmetry ${\hat \kappa}$ given by:
\be
\label{or_global}
([u_1,u_2], [v_1,v_2],z,u,x,y)\stackrel{\hat \kappa}{\longrightarrow}
([-v_2,v_1], [-u_2,u_1],-z,u,-y,-x) \ee which acts
as follows on the affine coordinates $\xi_j$ of the
two $\bfP^1$ factors: \be \xi_1\longleftrightarrow -1/\xi_2~~ \ee and
projects to the following involution $\kappa_0$ of $X_0$: \be
\label{or_projected}
(z,~x,~y,~u)\stackrel{\kappa_0}{\longrightarrow} (-z,~-y,~-x,~u)~~.  \ee

\subsection{The blown-up conifold case}

We now use  two $\bfP^1$ cycles but in the blown-up conifold.
In this situation we have
 \be
\label{rho_globalblowup}
([u_1,u_2], [v_1,v_2],x_1,x_2,x_3,x_4)\stackrel{{\hat \rho}(\theta)}{\longrightarrow}
([e^{-i\theta}u_1,u_2],
[v_1,e^{i\theta}v_2],x_1,x_2,e^{i\theta}x_3,e^{-i\theta} x_4)~~.  \ee
This projects as follows on the singular space $X_0$:
\be
\label{U1projblowup}
(x_1,~x_2,~x_3,~x_4)\stackrel{\rho_0(\theta)}{\longrightarrow} (x_1,~x_2,~e^{i\theta}x_3,~e^{-i\theta}x_4)~~.
\ee
The $Z_2$ symmetry ${\hat \kappa}$ is given in the homogeneous coordinates
$(u^1,u^2,v^1,v^2,s)$ by:
\bg
\label{or_globalblowup}
(u^1,u^2,v^1,v^2,s)\stackrel{\hat\kappa}{\longrightarrow}(-v^2,v^1,-u^2,u^1,s)
\nd
%\be
%\label{or_global}
%([u_1,u_2], [v_1,v_2],x_1,x_2,x_3,x_4)\stackrel{\hat \kappa}{\longrightarrow}
%([-v_2,v_1], [-u_2,u_1],-x_2,x_1,-x_4,-x_3) \ee
which projects on $X_0$ as:
% \be
%(x_1,x_2,x_3,x_4) \rightarrow (-x_1,-x_3,-x_2,x_4)
%\ee
\bg
(x_1,x_2,x_3,x_4) \rightarrow (x_2,x_1,-x_3,-x_4)
\nd
as is seen from (\ref{bdtoconifold}).

The action on the homogeneous coordinates $y_i$ of $\bfP^3$ is then given by
\be
(y_1,y_2,y_3,y_4) \rightarrow (y_2,y_1,-y_3,-y_4)
\ee
and the action on $\Omega$ is $\Omega \rightarrow - \Omega$.

What is the difference between the orientifold of \cite{orientifold} and the current one? The orientifold of
\cite{orientifold} was an O5 orientifold extended on directions orthogonal to the D5 branes wrapped on
$\bfP^1$ cycle. After the T-duality, the involution determined an inversion of the radial direction of the
$S^1$ which implied that the orientifold  became an O6 plane.

In the current example, the orientifold extends along the direction $s$ of the complex line bundle $L$ but also needs to
as an inversion on the ${\bf P^1} \times {\bf P^1}$ fiber which means that it wraps a 2-dimensional surface in the
${\bf P^1} \times {\bf P^1}$ fiber. Therefore we have O7 planes with the action as before.

\section{Analysis of the global picture and the cycle of geometric transitions}

With all the mathematical construction at hand, it is time now to discuss the geometrical aspect of the problem i.e
the supergravity metric and the fluxes in type II theories. Our starting point would be the
issue of supersymmetry in the usual resolved conifold background with fluxes and branes in type IIB background. Once
we obtain this, it will prepare us for all the subsequent stages of the duality cycle for the
geometric transition \cite{gtpapers}.

\subsection{Analysis of the global picture in type IIB}

{}From our earlier works we know that there are two ways of extending our local configuration of \cite{gtpapers} to study
supersymmetric cases in the full global picture:

\noindent (a) The full global geometry is a six-dimensional K\"ahler manifold with F-theory
seven-branes distributed in some particular way. These seven-branes contribute to massive
fundamental flavors in the gauge theory. Orientation of these seven-branes are the generalised version of the
Ouyang \cite{ouyang}
(or the Kuperstein \cite{kuperstein}) embeddings.

\noindent (b) The full global geometry is non-K\"ahler with or without F-theory seven-branes. The seven-branes could
be embedded in this picture via Ouyang or the Kuperstein embedding, which in turn would provide fundamental matters
in the gauge theory. In fact the possibility of such a global completion was already hinted in the second paper of
\cite{gtpapers}.

Let us see how from our local picture studied in \cite{gtpapers} these two possibilities can be realised. In the
first paper of \cite{gtpapers}, the local metric was argued to be of the following 
form\footnote{The local metric \eqref{localmet}
that we consider here is that of
a supergravity background studied around a specific chosen point in the internal six-dimensional space. For 
example we choose a point 
($r_0, \langle\theta_i\rangle, \langle\phi_i\rangle, \langle\psi\rangle$) in 
\cite{gtpapers} which is away from the $r=0$ conifold point. 
 This is because the full global picture was
hard to construct, and any naive procedure always lead to non-supersymmetric
solutions. In deriving the local metric, we took a simpler model where all
the spheres were replaced by tori with periodic coordinates ($x,
\theta_1$) and ($y,\theta_2$). The coordinate $z$ formed a
non-trivial $U(1)$ fibration over the $T^2$ base. Here ($r, x, y, z, \theta_1, \theta_2$) is the coordinate of a point 
away from 
($r_0, \langle\phi_1\rangle, \langle\phi_2\rangle, \langle\psi\rangle, \langle\theta_1\rangle, 
\langle\theta_2\rangle$). 
The replacement
of spheres by two tori was directly motivated from the
corresponding brane constructions of \cite{dasmukhi}, where
non-compact NS5 branes required the existence of tori instead of
spheres in the T-dual picture. On the other hand 
the term {\it global} means roughly adding back the curvature, warping, etc., replacing tori by
spheres, so that at the end of the day, we have a supersymmetric
solution to the equations of motion. The purpose of this paper 
is to exactly fill in the long-awaited gap, i.e to provide the full global picture of geometric
transition. Note also that the only known global solution, i.e \cite{pandoz}, before our work
was unfortunately not supersymmetric (see
\cite{cvetic}, \cite{anke} for details) although it satisfied the type IIB EOMs.}:
\bg\label{localmet}
ds^2 ~=~ && dr^2 + \Bigg(dz + \sqrt{\gamma' \over \gamma}~r_0 ~{\rm
cot}~\langle\theta_1\rangle~ dx  + \sqrt{\gamma' \over (\gamma+4a^2)~}~r_0~ {\rm
cot}~\langle\theta_2\rangle~dy\Bigg)^2~ + \nonumber\\
&& ~~~~~~~~~~~ + \Bigg[{\gamma\sqrt{h} \over 4}~d\theta_1^2 + dx^2\Bigg] +
\Bigg[{(\gamma + a^2)\sqrt{h} \over 4}~d\theta_2^2 + dy^2\Bigg] + ....
\nd
where all the coefficients are measured at a fixed chosen point ($r_0, \langle\psi\rangle, \langle\phi_i\rangle,
\langle\theta_i\rangle$). For more details see \cite{gtpapers}. The local $B_{\rm NS}$ field was taken to be:
\bg\label{bnslocal}
B_{\rm NS} = b_{x\theta_i} dx \wedge d\theta_i + b_{y\theta_i} dy \wedge d\theta_i
\nd
where $i= 1,2$. The above background is invariant under the orbifold operation:
\bg\label{orbi} {\cal I}_{xy}: ~~ x ~ \to ~-x, ~~~~ y ~\to ~ -y
\nd
and therefore can support D7/O7 states at the following orientifold points:
\bg\label{opoitla}
{{\bf T}^2 \over {\cal I}_{xy}~ \Omega ~(-1)^{F_L}}
\nd
It is interesting to note that, at the orientifold point, a component like $b_{xy}$ is projected out. However the
orientifold projection may allow components like $b_{xz}, b_{yz}$ which could in principle make our mirror manifold
non-geometric. In the local picture advocated in \cite{gtpapers} we only see components like \eqref{bnslocal} so
the local mirror is non-K\"ahler and geometric.

More interestingly, the orientifolding operation \eqref{opoitla} allows, along with the wrapped D5-branes,
the D7-branes and O7-planes along the internal directions ($r, z, \theta_1, \theta_2$) located at the
four fixed points of the torus ${\bf T^2}$ along ($x, y$) directions. Therefore,
in the local picture, a possible
susy preserving Ouyang-type configuration would be D5-branes wrapped on the two-torus ($\theta_2, \phi_2$) and the
seven branes wrapping ($\theta_1, \theta_2, \psi$) and stretched along the radial direction $r$. On the other hand,
globally
in a resolved conifold background the seven-branes are in a configuration that is the {\it union} of
branch 1 and branch 2 (see \cite{ouyang, sullyj, FEP, jpsi} for details). Recall that
in branch 1 the seven-branes wrap the ${\bf P^1}$ parametrised by ($\theta_2, \phi_2$) and are
embedded along ($r, \psi$) directions at a point on the other ${\bf P^1}$ parametrised by ($\theta_1, \phi_1$);
whereas in branch 2 the
seven-branes are at a point on the ${\bf P^1}$
parametrised by ($\theta_2, \phi_2$). Thus globally a susy configuration of
seven-branes is a {\it two}-dimensional surface in ${\bf P^1} \times {\bf P^1}$
and stretched along ($r, \psi$) directions determined
by the appropriate embedding equation. Therefore in the local limit the two-dimensional susy preserving surface
in ${\bf T^2} \times {\bf T^2}$ should be the two-cycle parametrised by ($\theta_1, \theta_2$) as
prescribed in \cite{gtpapers}.

Away from the orientifold point, the local metric takes the following fibration form:
\bg\label{locfib}
ds^2 = && ~h^{-1/2}ds^2_{0123} +  \gamma'\sqrt{h}~ dr^2 +
(dz + \Delta_1~{\rm cot}~\theta_1~ dx + \Delta_2~{\rm cot}~\theta_2~
dy)^2 + \nonumber\\
&& ~~~+ \left({\gamma \sqrt{h} \over 4} d\theta_1^2 + dx^2\right)
 + \left({(\gamma + 4a^2)\sqrt{h} \over 4} d\theta_2^2 + dy^2 \right)\nonumber\\
H_3 = && d{\cal J}_1 \wedge d\theta_1 \wedge dx + d{\cal J}_2 \wedge
d\theta_2 \wedge dy \nonumber\\
F_5 = && K(r)~(1 + \ast) ~dx \wedge dy \wedge
dz \wedge d\theta_1\wedge d\theta_2\\
F_3 = && c_1~(dz \wedge d\theta_2 \wedge dy - dz \wedge d\theta_1 \wedge dx)\nonumber
\nd
with additional axio-dilaton that appear from the seven-brane sources. The above form of orientifold projection only
allows a non-trivial fibration structure {\it away} from the orientifold point. However there exist another orientifold
operation that may be more well suited at the orientifold point. This can be applied locally via:
\bg\label{orbnow}
{\cal I}_{x\theta_1}: ~~~ x ~\to ~ -x, ~~~~~~~~~ \theta_1 ~\to ~ \pi - \theta_1
\nd
The above action gives rise to the following orientifold action:
\bg\label{oriennnw}
{{\bf T}^2 \over {\cal I}_{x\theta_1}~ \Omega ~(-1)^{F_L}}
\nd
that will keep the wrapped D5 branes only if they
are {\it away} from the orientifold point unless of course
they exist as bound states with the seven-branes {\it at} the orientifold point.
In addition there is the $B_{\rm NS}$ field with the following components at the orientifold point:
\bg\label{hecchu}
B_{\rm NS} ~=~&& b_{x\theta_2}~ dx \wedge d\theta_2 + b_{y\theta_1}~ dy \wedge d\theta_1 +
b_{xy}~ dx \wedge dy + b_{xz}~ dx \wedge dz + \nonumber\\
&& b_{rx}~ dr \wedge dx + b_{r\theta_1}~ dr \wedge d\theta_1 + b_{\theta_1\theta_2}~ d\theta_1 \wedge d\theta_2
+ b_{z\theta_1}~ dz \wedge d\theta_1
\nd
which means that at the orientifold point not only is the IIB metric non-trivial, the mirror can also be
non-K\"ahler and non-geometric. The seven-branes and the orientifold-planes are parallel to the wrapped D5-brane
bound states\footnote{These are in fact the dipole-deformed bound states studied in the last paper of
\cite{gtpapers}.}.
In the following we will argue how susy is preserved in the global set-up when the seven-branes are moved away from the
wrapped D5 branes. This is the case where the fundamental hypermultiplets are infinitely massive and therefore susy
remains unbroken at the scale that we want to study.

The naive global extension of the above configuration along the lines of \cite{pandoz} will lead to a non-susy
configuration. This is because we have assumed that the global extension of a configuration like \eqref{locfib} is
K\"ahler in the presence of a $B_{\rm NS}$ field like \eqref{bnslocal} {\it away} from the orientifold point. The
simplest global extension that we will study here as the starting point for the IIB geometric transition is a
non-K\"ahler manifold with D5-branes wrapping two cycles of this manifold. Of course it may be possible to add other
branes and fluxes to make the ambient space conformally K\"ahler, but we will not do so here.
We will use the following set of duality transformations,
recently proposed by \cite{marmal} (see also figure 1), to get
our type IIB intial configuration.

\noindent $\bullet$ Our starting point would be a non-K\"ahler type IIB metric with a background dilaton $\phi$ and
NS three-form $H_3$ that satisfies the standard relation $H_3 = e^{2\phi} \ast d(e^{-2\phi}J)$
with $J$ being the fundamental
(1,1) form.

\noindent $\bullet$ On this background we perform a S-duality that transforms the NS three-form to RR three-form
$F_3$, and in the process converts the dilaton to $-\phi$ without changing the metric in the Einstein frame.

\noindent $\bullet$ We now make three T-dualities along the spacetime directions $x^{1, 2, 3}$ that takes us to type
IIA theory. Observe that this is {\it not} the mirror construction.

\noindent $\bullet$ We lift the type IIA configuration to M-theory and perform a boost along the eleventh direction.
This boost is crucial in generating D0-brane {\it gauge} charges in M-theory.

\noindent $\bullet$ A dimensional reduction back to IIA theory does exactly what we wanted: it generates the necessary
number of D0-brane charges from the boost, without breaking the underlying supersymmetry of the system.

\noindent $\bullet$ Once we have the IIA configuration, we go back to type IIB by performing the three T-dualities
along $x^{1, 2, 3}$ directions. From the D0-branes, we get back our three-brane charges namely the five-form. The
duality cycle also gives us NS three-form $H_3$ as well as the expected RR three-form $F_3$. Therefore the final
configuration is exactly what we required for IR geometric transition: wrapped D5s with necessary sources on a
non-K\"ahler globally defined ``resolved'' conifold background. Also as expected, the background preserves
supersymmetry and therefore should be our starting point. This background should also be compared with the one
given in \cite{pandoz} that solves the EOM but does not preserve supersymmetry. 
In figure 1, we illustrate the above dualities.
\begin{figure}[htb]\label{MMdualities}
       \begin{center}
\includegraphics[height=12cm]{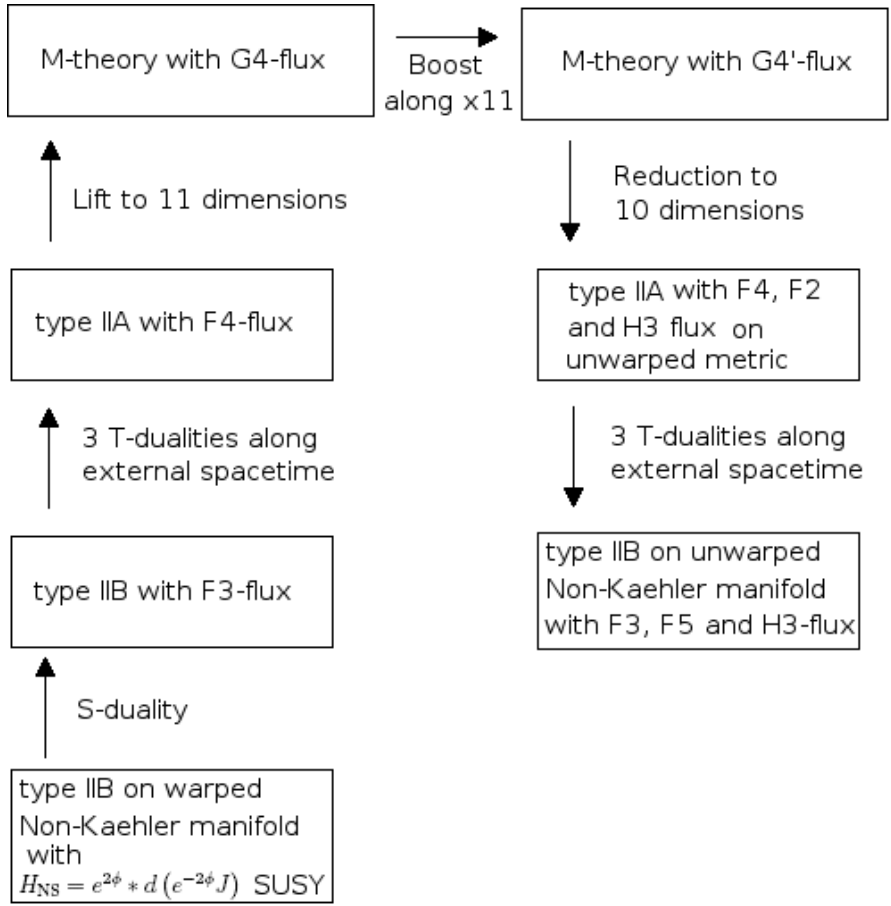}
        \caption{This figure illustrates the series of dualities that we used to generate
the full supersymmetric background with non-trivial fluxes.}
        \end{center}
        \end{figure}
To start off, we switch on a non-trivial background dilaton $\phi$ and a NS three-form $H_{\rm NS}$ on a
background outlined by the following metric:
\bg \label{startmet}
ds^2 = h^{1/2} e^\phi d{\widetilde s}^2_{0123} + h^{-1/2} e^\phi ds^2_6
\nd
where note that we can choose the dilaton $\phi$ appropriately so that in the string frame the spacetime metric may not
have a warp factor. This will be consistent with the last reference of \cite{torsion}.
We have defined the other variables in the following way:
\bg \label{varde}
&&h = {e^{-2\phi} F_0^{-4} \over e^{-2\phi} h^{-2}F_0^{-4}{\rm cosh}^2\beta - {\rm sinh}^2 \beta},
~~~~~~ d{\widetilde s}^2_{0123} = F_0 ds^2_{0123}\\
&& ds^2_6 = F_1~ dr^2 + F_2 (d\psi + {\rm cos}~\theta_1 d\phi_1 + {\rm cos}~\theta_2 d\phi_2)^2  + \sum_{i = 1}^2 F_{2+i}
(d\theta_i^2 + {\rm sin}^2\theta_i d\phi_i^2)\nonumber
\nd
with $\beta$ being an arbitrary constant, $F_i \equiv F_i(r), i = 1, ..., 4$ 
are functions of the radial coordinate for simplicity and $F_0 = F_0(r, \theta_1, \theta_2)$ since this is unconstrained. 
Observe that in the first equation of \eqref{varde}, $h$ appears on both sides, and once we put in the value of the
dilaton we can determine the warp factor $h$. For our case the dilaton will take the following form:
\bg\label{dildil}
\phi ~ = ~ -{\rm log}~F_0 - {1\over 2} {\rm log}~h
\nd
so that the starting metric in IIB, that preserves spacetime supersymmetry, becomes:
\bg\label{smetru}
ds^2 ~ = ~ ds^2_{0123} ~+~ \left({1+F_0^2 {\rm sinh}^2\beta\over F_0 {\rm cosh}^2\beta}\right) ds^2_6
\nd
In
general we will continue keeping the dilaton $\phi$ in the metric components to get the general torsion classes for the
background (see the analysis in \cite{chen2}). 
We also
expect $F_i$ to be functions of all the internal coordinates. We will give an example of this soon when
we derive a more precise initial metric. For the time being we will consider \eqref{startmet} to be our starting point.
Also to preserve supersymmetry\footnote{Or, equivalently, preserving $SU(3)$ structure.}, we expect:
\bg\label{hns}
H_{\rm NS} = e^{2\phi} \ast d\left(e^{-2\phi} J\right)
\nd
with the appropriate dilaton. Here
$J$ is the fundamental form associated with the metric, and we can choose to
impose one of the following two conditions on the
NS three-form:
\bg\label{condhns}
&&dH_{\rm NS} ~\equiv~ d\ast dJ - d\ast (d\phi \wedge J) ~=~ {\rm sources}\nonumber\\
&&dH_{\rm NS} ~\equiv~ d\ast dJ - d\ast (d\phi \wedge J) ~ = ~\alpha'({\rm tr} ~R \wedge R - {\rm Tr} ~ F \wedge F)
\nd
The first condition is what we require here. This will give rise to the IR wrapped D5 branes theory on non-K\"ahler
resolved conifold set-up (after the chain of dualities mentioned above). The latter case will be for the
heterotic theory. We can use the non-closure of $H_{\rm NS}$ to study not only
the vector bundles $F$ on the heterotic side, but also the possibility of geometric transition in the heterotic theory!
We have alluded to this possibility in our earlier papers \cite{gtpapers} (see also \cite{israel}).
We have completed that side of the
story in our follow-up paper \cite{chen2}.

Now following the chain of dualities mentioned above, we can get the following type IIB 
background\footnote{Due to an unfortunate choice of notation, the RR three-form and the third warp factor have 
the same notation of $F_3$ (as this is the standard way to represent them!). 
However since we use $F_3$ to mostly denote the third warp factor, we hope that there will be 
no confusion.}:
\bg\label{susybg}
&& F_3 = h~{\rm cosh}~\beta~e^{2\phi} \ast d\left(e^{-2\phi} J\right), ~~~~~~~
H_3 = -hF_0^2{\rm sinh}~\beta~e^{2\phi} d\left(e^{-2\phi} J\right)\nonumber\\
&& F_5 = -{1\over 4} (1 + \ast) dA_0 \wedge dx^0 \wedge dx^1 \wedge  dx^2 \wedge  dx^3, ~~~~~
\phi_{\rm now} = -\phi\\
&& ds^2 = F_0 ds^2_{0123} + F_1~ dr^2 + F_2 (d\psi + {\rm cos}~\theta_1 d\phi_1 + {\rm cos}~\theta_2 d\phi_2)^2
+ \sum_{i = 1}^2 F_{2+i}
(d\theta_i^2 + {\rm sin}^2\theta_i d\phi_i^2)\nonumber
\nd
which is, by construction, supersymmetric and since the RR three-form
$F_3$ is not closed it represents precisely the IR configuration
of wrapped D5-branes on warped non-K\"ahler resolved conifold. The above set of equations \eqref{susybg} is one 
of our main results, and as promised in the introduction, provides the fully supersymmetric globally defined 
solution for the wrapped D5-branes on a non-K\"ahler resolved conifold. 
Note that we have left the warp factors $F_i$ undetermined in \eqref{susybg}. This means that 
there is a wide range of choices for $F_i$ related to various gauge theory deformations. This is closely related to 
a similar class of solutions illustrated in figure 3 of \cite{chen2}. Thus the procedure will be to identify 
certain set of $\{F_i\}$ related to deformations in ${\cal N} = 1$ YM theory, and then our duality chain will reproduce 
the gravity dual of this YM configuration. One may also refer to a recent class of solutions studied in \cite{pando2} 
with a given choice of $\{F_i\}$. 

Finally, the five-form is switched on to satify the equation of
motion with
\bg\label{a0def}
A_0 ~&=~ &{{\rm cosh}~\beta~{\rm sinh}~\beta (1-e^{2\phi}h^{-2}F_0^{-4})
\over e^{-2\phi} h^{-2} F_0^{-4}{\rm cosh}^2\beta - {\rm sinh}^2\beta}\nonumber\\
&= ~& (F_0^2 -1 ){\rm tanh}~\beta \left[1 + \left({1-F_0^2\over F_0^2}\right){\rm sech}^2\beta +  
\left({1-F_0^2\over F_0^2}\right)^2 {\rm sech}^4 \beta\right]
\nd
The above equation \eqref{susybg} 
is our starting metric, whose local forms
we studied in details in \cite{gtpapers}, and therefore should be taken instead of the metric derived in \cite{pandoz}.
The ISD condition for our case gets modified to the following condition on the fluxes:
\bg\label{isdnow}
\cosh\beta~H_3 ~+~ F_0^2~\sinh\beta \ast F_3 ~ = ~ 0
\nd
which may be compared to \cite{marmal}.
Our derivation
could also solve the long standing problem of finding the supersymmetric configuration of wrapped D5-branes on a
resolved conifold set-up.

\subsection{More explicit type IIB background before geometric transition}

In the above section we saw how one could derive the precise intial
metric that not only serves as starting point for geometric
transition, but is also supersymmetric. One may make this more specific by
solving the $SU(3)$ structure condition specified in section 2.4. This is
worked out in {\bf Appendix 1}.
The metric derived this way has
many non-trivial components compared to our initial ansatze
\eqref{susybg}. This is not a problem in itself, because we can
always do some coordinate transformations to bring the metric to a form that
doesn't have components like $g_{r\mu}$ where $\mu = \theta_i,
\phi_i, \psi$. But the metric will have other non-trivial
cross-terms. It may be possible to make further coordinate
transformations to bring the above metric in a form that closely
resembles \eqref{susybg}, but we will not pursue this here as this
doesn't change the underlying physics. Instead we will continue
using the background \eqref{susybg} and assume that the values of
the coefficients $F_i$ are to be fixed using our above metric
configuration\footnote{See also \cite{israel} where a non-K\"ahler metric
on the resolved conifold is studied. It would be interesting to compare the
metric of \cite{israel} with the metric components given in {\bf Appendix 1}.}.
Other possible cross-terms, not considered in
\eqref{susybg}, will only make the IIA background more non-trivial
without revealing new physics\footnote{We have justified this claim in \cite{chen2} where we explicitly computed the 
torsion classes for the background \eqref{startmet}, \eqref{dildil} and \eqref{hns} to argue for supersymmetry. See 
section 3.1 of \cite{chen2}.}. 
Henceforth our starting point would
be \eqref{susybg} with the assumption that the coefficients are to
be derived from the metric discussed in the above subsection.
\begin{figure}[htb]\label{dualitiesGT}
        \begin{center}
\includegraphics[height=10cm]{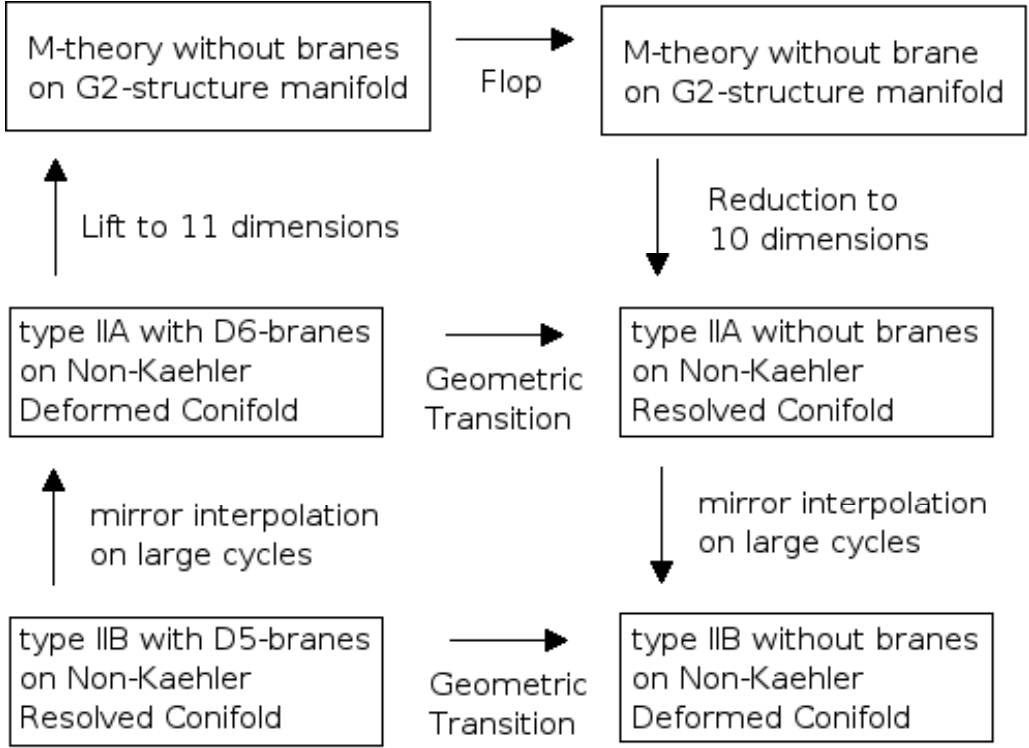}
        \caption{The duality map to generate the full geometric transitions in the supersymmetric global
set-up of type IIA and type IIB theories.}
        \end{center}
        \end{figure}
Once we know the metric, we can follow up the steps described
earlier to compute the three-forms. The NS three-form $H_3$ has the
form: \begin{eqnarray}\label{H3}
\frac{H_3}{hF_0^2\sinh\beta}=&+&\Big[k^2(2\phi_{\theta_1}\sqrt{F_1F_2}\cos\theta_1+\sqrt{F_1F_2}\sin\theta_1+2\phi_rF_3\sin\theta_1-F_{3r}\sin\theta_1)\nonumber\\
&-&2kF_3k_r\sin\theta_1-2k\sqrt{F_1F_2}k_{\theta_1}\cos\theta_1\Big]dr\wedge
d\theta_1\wedge d\phi_1\nonumber\\
&+&\Big[k^2(2\phi_{\theta_2}\sqrt{F_1F_2}\cos\theta_2+\sqrt{F_1F_2}\sin\theta_2+2\phi_rF_4\sin\theta_2-F_{4r}\sin\theta_2)\nonumber\\
&-&2kF_4k_r\sin\theta_2-2k\sqrt{F_1F_2}k_{\theta_2}\cos\theta_2\Big]dr\wedge
d\theta_2\wedge d\phi_2\nonumber\\
&-&2k\sqrt{F_1F_2}\Big[(k\phi_{\theta_1}-k_{\theta_1})(dr\wedge
d\psi\wedge d\theta_1-\cos\theta_2dr\wedge d\theta_1\wedge
d\phi_2)\nonumber\\
&+&(k\phi_{\theta_2}-k_{\theta_2} )(dr\wedge d\psi\wedge
d\theta_2-\cos\theta_1dr\wedge
d\theta_2\wedge d\phi_1)\Big]\nonumber\\
&-&2kF_3\sin\theta_1(k\phi_{\theta_2}-k_{\theta_2})d\theta_1\wedge
d\theta_2\wedge
d\phi_1\nonumber\\
&+&2kF_4\sin\theta_2(k\phi_{\theta_1}-k_{\theta_1})d\theta_1\wedge
d\theta_2\wedge d\phi_2
\end{eqnarray}
where $k^2(r,\theta_1,\theta_2)=h^{-1/2}e^{\phi}$, and we have
defined $\phi_\alpha \equiv \partial_\alpha \phi$ with $\alpha =
\theta_i, r$ as $\phi \equiv \phi (r, \theta_1, \theta_2)$ for
simplicity. A constant $\phi$ is not good for us, and also leads to
certain issues detailed in \cite{papad}. Once we have $H_3$, we can
get $dH_3$ as:
\begin{eqnarray}\label{dH3}
\frac{dH_3}{\sinh\beta}&=&k\Big[F_3\sin\theta_1(4hk_r\phi_{\theta_2}-4hk_{\theta_2}\phi_r+2h_{\theta_2}k_r-2h_rk_{\theta_2}+2kh_r\phi_{\theta_2}-2kh_{\theta_2}\phi_r)\nonumber\\
&&\sqrt{F_1F_2}\cos\theta_1(4hk_{\theta_1}\phi_{\theta_2}-4hk_{\theta_2}\phi_{\theta_1}+2h_{\theta_2}k_{\theta_1}-2h_{\theta_1}k_{\theta_2}
+2kh_{\theta_1}\phi_{\theta_2}-2kh_{\theta_2}\phi_{\theta_1})\nonumber\\
&&-(\sqrt{F_1F_2}-F_{3r})(2hk\phi_{\theta_2}+h_{\theta_2}k)\sin\theta_1\Big]dr\wedge
d\theta_2\wedge d\theta_1\wedge d\phi_1\nonumber\\
&&+k\Big[F_4\sin\theta_2(4hk_r\phi_{\theta_1}-4hk_{\theta_1}\phi_r+2h_{\theta_1}k_r-2h_rk_{\theta_1}+2kh_r\phi_{\theta_1}-2kh_{\theta_1}\phi_r)\nonumber\\
&&\sqrt{F_1F_2}\cos\theta_2(4hk_{\theta_2}\phi_{\theta_1}-4hk_{\theta_1}\phi_{\theta_2}+2h_{\theta_1}k_{\theta_2}-2h_{\theta_2}k_{\theta_1}
+2kh_{\theta_2}\phi_{\theta_1}-2kh_{\theta_1}\phi_{\theta_2})\nonumber\\
&&-(\sqrt{F_1F_2}-F_{4r})(2hk\phi_{\theta_1}+h_{\theta_1}k)\sin\theta_2\Big]dr\wedge
d\theta_1\wedge d\theta_2\wedge
d\phi_2\nonumber\\
&&+2k\sqrt{F_1F_2}\Big[h_{\theta_1}k_{\theta_2}-h_{\theta_2}k_{\theta_1}+kh_{\theta_2}\phi_{\theta_1}-kh_{\theta_1}\phi_{\theta_2}+2hk_{\theta_2}\phi_{\theta_1}-2hk_{\theta_1}\phi_{\theta_2}\Big]\nonumber\\
&&~~~~~~~~~~~~~~~~\times dr\wedge d\theta_1\wedge d\theta_2\wedge
d\psi
\end{eqnarray}
with $F_{ir} \equiv \partial_r F_i$, $k_{i}=\partial_i k$ and
$(hF_0^2)_{i}=\partial_i (hF_0^2)$. From \eqref{dH3} it means that
if we make
\begin{eqnarray}
&&h_{\theta_1}k_{\theta_2}-h_{\theta_2}k_{\theta_1}+kh_{\theta_2}\phi_{\theta_1}-kh_{\theta_1}\phi_{\theta_2}+2hk_{\theta_2}\phi_{\theta_1}-2hk_{\theta_1}\phi_{\theta_2}=0,\nonumber\\
&&\frac{4hk_r\phi_{\theta_2}-4hk_{\theta_2}\phi_r+2h_{\theta_2}k_r-2h_rk_{\theta_2}+2kh_r\phi_{\theta_2}-2kh_{\theta_2}\phi_r}{2hk\phi_{\theta_2}+h_{\theta_2}k}=\frac{\sqrt{F_1F_2}-F_{3r}}{F_3},\nonumber\\
&&\frac{4hk_r\phi_{\theta_1}-4hk_{\theta_1}\phi_r+2h_{\theta_1}k_r-2h_rk_{\theta_1}+2kh_r\phi_{\theta_1}-2kh_{\theta_1}\phi_r}{2hk\phi_{\theta_1}+h_{\theta_1}k}=\frac{\sqrt{F_1F_2}-F_{4r}}{F_4}.\nonumber
\end{eqnarray}
then $H_3$ will be closed. One however may worry that making $H_3$
closed implies too much constraints on the $F_i$'s. For the present
case this may still be okay because the initial choice of the
background \eqref{varde} forms a class of solutions parametrised by
our choice of $F_i$ and $\phi$, the dilaton\footnote{See also \cite{chen2} where a torsion class analysis reveals 
$k$ to be a function of $r$ only. This means we can choose $F_0$ appropriately such that $h$ and $e^\phi$ are 
functions of ($r, \theta_i$). One possible choice would be $h = {1\over k^2 F_0}$ and $e^\phi = {k\over \sqrt{F_0}}$.}. 
A specific choice of the
background with a specified complex structure and K\"ahler class is
exemplified in {\bf Appendix 1}. For this case we can define a
closed three-form with appropriate choice of the dilaton, so that
our choice remains generic enough. Thus the $B_{NS}$ field can be
gauge transformed to have only the following components:
\begin{eqnarray}\label{bns}
b_{r\psi}&=&\int-2hF_0^2k\sinh\beta~(k\phi_{\theta_1}-k_{\theta_1})\sqrt{F_1F_2}~d\theta_1,
\nonumber\\
b_{r\phi_1}&=&\int-2hF_0^2k\sinh\beta~(k\phi_{\theta_2}-k_{\theta_2})\sqrt{F_1F_2}\cos\theta_1~d\theta_2,
\nonumber\\
b_{r\phi_2}&=&\int-2hF_0^2k\sinh\beta~(k\phi_{\theta_1}-k_{\theta_1})\sqrt{F_1F_2}\cos\theta_2~d\theta_1,\nonumber\\
b_{\theta_1\phi_1}&=&\int2hF_0^2k\sinh\beta~(k\phi_{\theta_2}-k_{\theta_2})
F_3\sin\theta_1~d\theta_2,\nonumber\\
b_{\theta_2\phi_2}&=&\int2hF_0^2k\sinh\beta~(k\phi_{\theta_1}-k_{\theta_1})
F_4\sin\theta_2~d\theta_1.
\end{eqnarray}
where we see that there are three new components of the form $b_{r\alpha}$ compared to the local case \cite{gtpapers}.
This is expected because we are no longer fixed to $r = r_0$, but have global access. However
before moving ahead we will pause to comment on switching on
other possible components of the $B_{NS}$ field of the form:
\bg\label{bnsother}
 b_{\phi_1 \phi_2}~ d\phi_1 \wedge d\phi_2 ~+~  \sum_{i = 1}^2 b_{\phi_i \psi}~ d\phi_i \wedge d\psi
\nd
Such choices of $B_{NS}$ fields would make the type IIA background {\it non-geometric}. So far locally we saw that the
type IIA backgrounds remains geometric but does become non-K\"ahler. Is there a possibility that the IIA background
globally is non-geometric also? We will reflect on this point later, but for the time being we will assume that the
$B_{NS}$ field is only of the form \eqref{bns} and doesn't have additional components like \eqref{bnsother}.

Next comes the RR three-form $F_3$. From our previous set of duality arguments, this is given by:
\begin{eqnarray}\label{F3}
\frac{F_3}{h
F_0^2\cosh\beta}=&&2KF_1F_2F_3F_4\sin\theta_2\sin\theta_1(\phi_{\theta_1}\sin\theta_1\cos\theta_2-
\phi_{\theta_2}\sin\theta_2\cos\theta_1)dr \wedge d\phi_1\wedge d\phi_2\nonumber\\
&&+KF_3^2\sin^2\theta_1\sin\theta_2(2\phi_{\theta_2}\sqrt{F_1F_2}F_4\sin\theta_2-F_2\sqrt{F_1F_2}
\cos\theta_2\nonumber\\
&&-2\phi_rF_2F_4\cos\theta_2+F_2F_{4r}\cos\theta_2)d\theta_1\wedge
d\phi_1\wedge d\phi_2\nonumber\\
&&+KF_4^2\sin^2\theta_2\sin\theta_1
(-2\phi_{\theta_1}\sqrt{F_1F_2}\sin\theta_1+F_2\sqrt{F_1F_2}\cos\theta_1\nonumber\\
&&+2\phi_rF_2F_3\cos\theta_1-F_2F_{3r}\cos\theta_1)d\theta_2\wedge d\phi_1\wedge d\phi_2\nonumber\\
&&-KF_2F_3^2\sin^2\theta_1(2\phi_rF_4\sin\theta_2+\sqrt{F_1F_2}\sin\theta_2-F_{4r}\sin\theta_2)d\psi\wedge
d\theta_1\wedge d\phi_1\nonumber\\
&&-KF_2F_4^2\sin^2\theta_2(2\phi_rF_3\sin\theta_1+\sqrt{F_1F_2}\sin\theta_1-F_{3r}\sin\theta_1)d\psi\wedge
d\theta_2\wedge d\phi_2\nonumber\\
&&-2\phi_{\theta_2}KF_1F_2F_3F_4\sin\theta_1\sin^2\theta_2dr\wedge
d\psi\wedge d\phi_2\nonumber\\
&&-2\phi_{\theta_1}KF_1F_2F_3F_4\sin\theta_2\sin^2\theta_1dr\wedge
d\psi\wedge d\phi_1
\end{eqnarray}
where as before $\phi_\alpha$ should be understood as derivatives on $\phi$ i.e $\partial_\alpha \phi$,
and we have defined $K$ as:
\bg\label{kdef}
K ={{\rm cosec}~\theta_1{\rm cosec}~\theta_2\over \sqrt{F_1F_2}F_3F_4}
\nd
Note that $dF_3$ is no longer
closed, and will be related to delta function sources coming from the wrapped 
D5-branes\footnote{There is a subtlety 
here: not every non-closed $F_3$ can be interpreted as a source (see for example the criteria presented in 
\cite{konkal}). Happily,
our case does fall into one of the required criteria of \cite{konkal} as should be clear by writing the 
fluxes in the language of G-structure, or in terms of the torsion classes. A more detailed elaboration of this is 
given in \cite{chen2}.}.

Once we have the explicit forms for the three-forms, to satisy the type IIB EOMs we will now require RR five-form.
This is easy to work out, and is given by:
\begin{eqnarray}\label{F5}
F_5&=&\frac{1}{4}\Big[-A_{0r}dr\wedge dt\wedge dx\wedge dy\wedge
dz-A_{0\theta_1}d\theta_1\wedge dt\wedge dx\wedge dy\wedge
dz\nonumber\\
&&-A_{0\theta_2}d\theta_2\wedge dt\wedge dx\wedge dy\wedge dz
-PF_2F_3F_4\sin^2\theta_1\sin^2\theta_2\nonumber\\
&&\times (A_{0r}F_3F_4d\psi\wedge d\theta_1\wedge d\theta_2\wedge
d\phi_1\wedge d\phi_2+A_{0\theta_1}F_1F_4dr\wedge d\psi\wedge
d\theta_2\wedge
d\phi_1\wedge d\phi_2\nonumber\\
&& +A_{0\theta_2}F_1F_3dr\wedge d\psi\wedge d\theta_1\wedge
d\phi_1\wedge d\phi_2)\Big]
\end{eqnarray}
where $A_{0\alpha} \equiv \partial_\alpha A_0$ and $A_0$ is given in \eqref{a0def}. We have also defined $P$ as:
\bg\label{pdef}
P={{\rm cosec}~\theta_1{\rm cosec}~\theta_2\over \sqrt{F_1F_2}F_0^2F_3 F_4}
\nd
Thus with \eqref{H3}, \eqref{F3}, \eqref{F5}
and \eqref{susybg} we have the complete susy background in type IIB before geometric transition. A torsion
class analysis with susy constraints has been done in \cite{chen2} (see eq. (3.15) therein). As long as the
warp factors satisfy eq. (3.15) of \cite{chen2} supersymmetry will be preserved.
In the following subsection, we will use the above background to
compute the type IIA mirror configuration.

\subsection{The type IIA mirror configuration}

As it stands, the metric in \eqref{susybg} has three obvious
isometries associated with translation along the three angular
directions $\phi_1, \phi_2$ and $\psi$. So there is a natural ${\bf
T}^3$ embedded in our configuration, and one might naively think
that the mirror would be three T-dualities along ${\bf T}^3$. Such a
simple transformation doesn't work for our case because our
configuration represents the IR limit of a cascading gauge theory
where the base of the three torus is {\it small}. Mirror
transformation {\it a la} SYZ \cite{syz} works exactly in the
opposite limit! So naive T-dualities will not give us the mirror
metric, and we need to first make the base, paramerised by
$\theta_1, \theta_2$ and $r$, very large\footnote{This effectively
means that the distances along the $\theta_i$ directions have to be
made very large, as $r$ is non-compact. See also our earlier works \cite{gtpapers} where this
has been explained in more details.}. The simplest way to do
this would be to make the following transformation on the background
\eqref{susybg}: \bg\label{tranbg}
&& d\psi ~ \to ~ d\psi ~+~f_1~ {\rm cos}~\theta_1~d\theta_1   ~+~f_2~ {\rm cos}~\theta_2~d\theta_2\nonumber\\
&& d\phi_1 ~ \to ~ d\phi_1 ~ - ~ f_1~d\theta_1, ~~~~~~~ d\phi_2 ~
\to ~ d\phi_2 ~ - ~ f_2~d\theta_2 \nd with the assumption that $f_i
= f_i(\theta_i)$ so that the transformations \eqref{tranbg} would be
integrable\footnote{Note also that since $f_i = f_i(\theta_i)$, the transformations \eqref{tranbg} on the 
one-forms ($d\psi, d\phi_i$) are just coordinate transformations of ($\psi, \phi_i$). Therefore they don't change the 
EOMs.}.  
Recall that compatibility with the $SU(3)$ structure will fix $f_i$ in the mirror \cite{gtpapers}.
Note also that these transformations are similar in form as
in the first reference of \cite{gtpapers} and would change the complex structure of the base accordingly.

Under these transformations the $B_{\rm NS}$ field generates extra
components $b_{r\theta_1}$, $b_{r\theta_2}$. It is however interesting to note that
they vanish as follows:
\begin{eqnarray}
b_{r\theta_1}=f_1(b_{r\psi}\cos\theta_1-b_{r\phi_1})=0,\;\;b_{r\theta_2}=f_2(b_{r\psi}\cos\theta_2-b_{r\phi_2})=0
\end{eqnarray}
implying that the $B_{\rm NS}$ field do not change under the transformation \eqref{tranbg}. This is similar to the
local case also \cite{gtpapers}.

On the other hand the RR three-form {\it does} change under the coordinate transformation \eqref{tranbg}. The
extra components of the three-form are the following:
\begin{eqnarray}\label{extra3}
&&F_{r\theta_1\theta_2}=f_1f_2(F_{r\phi_1\phi_2}-\cos\theta_1F_{r\psi\phi_2}+\cos\theta_2F_{r\psi\phi_1}),
~~~F_{r\psi\theta_1}=-f_1F_{r\psi\phi_1}\nonumber\\
&&F_{r\theta_1\phi_2}=-f_1(F_{r\phi_1\phi_2}-\cos\theta_1F_{r\psi\phi_2}),\quad\,
F_{\theta_1\theta_2\phi_1}=f_2(F_{\theta_1\phi_1\phi_2}-\cos\theta_2F_{\psi\theta_1\phi_1}),
\nonumber\\
&&F_{r\theta_2\phi_1}=f_2(F_{r\phi_1\phi_2}+\cos\theta_2F_{r\psi\phi_1}),\quad\,
F_{\theta_1\theta_2\phi_2}=f_1(F_{\theta_2\phi_1\phi_2}+\cos\theta_1F_{\psi\theta_2\phi_2}),
\nonumber\\
&&F_{r\theta_2\phi_2}=f_2\cos\theta_2F_{r\psi\phi_2},~~~
F_{r\psi\theta_2}=-f_2F_{r\psi\phi_2},~~~
F_{r\theta_1\phi_1}=f_1\cos\theta_1F_{r\psi\phi_1}
\end{eqnarray}
A physical reason for this change can be easily understood: under the coordinate transformation \eqref{tranbg} the
base parametrised by ($\theta_1, \theta_2$) become large. This means that the associated RR three-form
field strengths increase
simultaneously, which is of course what we see in \eqref{extra3}. Note that the component
$F_{r\theta_1\theta_2}$ dominates over all other extra components in \eqref{extra3} because this lies exclusively
on the base parametrised by the coordinates ($r, \theta_1, \theta_2$) which is made much bigger than the
${\bf T}^3$ fibre parametrised by the coordinates ($\psi, \phi_1, \phi_2$).

Once the three-form $F_3$ changes, the RR five-form also has to change. Its is easy to show that the
extra components of the five-form are:
\begin{eqnarray}\label{extra5}
&&F_{r\theta_1\theta_2\phi_1\phi_2}=f_1\cos\theta_1F_{r\psi\theta_2\phi_1\phi_2}-f_2\cos\theta_2
F_{r\psi\theta_1\phi_1\phi_2},\nonumber\\
&&F_{r\psi\theta_1\theta_2\phi_2}=f_1F_{r\psi\theta_2\phi_1\phi_2},\quad
F_{r\psi\theta_1\theta_2\phi_1}=f_2F_{r\psi\theta_1\phi_1\phi_2}
\end{eqnarray}
satisfying the background EOMs. All these extra components will give rise to RR four-form in Type IIA
after mirror transformation, as we will show soon. But before that lets infer how the metric changes.
Under the transformation \eqref{tranbg} the metric \eqref{susybg} takes the
following form:
\bg\label{sysymet} ds^2 = && F_0 ds^2_{0123} + F_1~
dr^2 + F_2 (d\psi + {\rm cos}~\theta_1 d\phi_1 + {\rm cos}~\theta_2
d\phi_2)^2
+ \sum_{i = 1}^2 F_{2+i} ~{\rm sin}^2\theta_i d\phi_i^2 \nonumber\\
&& ~~~~~~~~~~~+ \sum_{i = 1}^2 \Big[F_{2+i}\left(1 + f_i^2~{\rm sin}^2 \theta_i\right) d\theta_i^2 -2 f_i F_{2+i}~
{\rm sin}^2 \theta_i ~d\phi_i d\theta_i\Big]
\nd
which in fact does exactly what we wanted\footnote{The metric \eqref{sysymet} also solves the supergravity EOM as 
should be clear from the discussion presented earlier.}: 
it enlarges the $\theta_i$-cycles, but doesn't change the $B_{\rm NS}$
field. For SYZ to work properly, we require the base size to be very large, and therefore we will require $f_i$ also to be
large. This conclusion fits well with the local picture that we had in \cite{gtpapers}. Note that we have also
generated cross terms. These cross terms will be useful soon. The eleven metric components are:
\bg\label{metcom}
&& j_{rr} = F_1, ~~~~ j_{\phi_1\theta_1} = -f_1F_3 {\rm sin}^2\theta_1, ~~~~
j_{\phi_2\theta_2} = -f_2 F_4 {\rm sin}^2\theta_2 \nonumber\\
&& j_{\psi\psi} = F_2(1-\epsilon), ~~~~ j_{\phi_1\psi} = F_2 {\rm cos}~\theta_1, ~~~~
j_{\phi_2\psi} = F_2 {\rm cos}~\theta_2\nonumber\\
&& j_{\phi_1\phi_1} = F_2 {\rm cos}^2 \theta_1 + F_3 {\rm sin}^2 \theta_1, ~~~~
j_{\phi_2\phi_2} = F_2 {\rm cos}^2 \theta_2 + F_4 {\rm sin}^2 \theta_2\\
&& j_{\phi_1\phi_2} = F_2 {\rm cos}~\theta_1 \theta_2, ~~~~
j_{\theta_1 \theta_1} = F_3 (1+f_1^2 {\rm sin}^2 \theta_1), ~~~
j_{\theta_2 \theta_2} = F_{4} (1+f_2^2 {\rm sin}^2
\theta_2)\nonumber \nd where $\epsilon$ is a finite but small number\footnote{Of course $\epsilon < 1$ to preserve the 
signature. 
In fact introducing $\epsilon$ in
$j_{\psi\psi}$ will help us not only to keep $f_i$ large to satisfy SYZ but also satisfy the susy conditions
in the mirror dual. This will become clear soon.}.
Let
us also define another quantity $\alpha$ in the following way:
\bg\label{alphadef} \alpha^{-1} ~\equiv~ {F_3 F_4 {\rm
sin}^2\theta_1 {\rm sin}^2\theta_2 + F_2 F_4 {\rm cos}^2\theta_1{\rm
sin}^2\theta_2 + F_2 F_3 {\rm sin}^2\theta_1 {\rm cos}^2\theta_2}
\nd away from the point ($\theta_1, \theta_2$) $=0$. Now assuming
that $f_1, f_2$ are large, we can perform the mirror
transformation along ($\psi, \phi_1, \phi_2$) directions. The mirror metric in type IIA takes the following
form: \bg\label{mirrormet} ds^2_{\rm mirror} = F_0 ds^2_{0123} +
ds^2_6 \nd where the six-dimensional internal space is a
non-K\"ahler deformation of the deformed conifold in the following
way: \bg\label{nkdefco} &&ds^2_6 = F_1 dr^2 + {\alpha F_2 \over
\Delta_1 \Delta_2} \Big[d\psi-b_{\psi r}dr + \Delta_1 {\rm
cos}~\theta_1 \Big(d\phi_1 - b_{\phi_1\theta_1} d\theta_1-b_{\phi_1
r}dr\Big)\nonumber\\
&&\quad\quad\quad\quad\quad\quad\quad\quad\quad\quad+ \Delta_2
{\rm cos}~\theta_2 \Big(d\phi_2 - b_{\phi_2\theta_2} d\theta_2-b_{\phi_2 r}dr\Big)\Big]^2\nonumber\\
&& ~~~~~~ + \alpha j_{\phi_2\phi_2}\Big(d\phi_1 - b_{\phi_1\theta_1}
d\theta_1-b_{\phi_1 r}dr\Big)^2 + \alpha
j_{\phi_1\phi_1}\Big(d\phi_2 - b_{\phi_2\theta_2}
d\theta_2-b_{\phi_2
r}dr\Big)^2\nonumber\\
&& ~~~~~~ -2\alpha j_{\phi_1\phi_2} \Big(d\phi_1 -
b_{\phi_1\theta_1} d\theta_1-b_{\phi_1 r}dr\Big)\Big(d\phi_2 -
b_{\phi_2\theta_2} d\theta_2-b_{\phi_2
r}dr\Big)\\
&& ~~~~~~ -2\alpha j_{\phi_1\phi_2} \left({f_1 f_2 \epsilon\over
\alpha}\right) d\theta_1 d\theta_2 + \Big(F_3 - \epsilon ~F_2 f_1^2
{\rm cos}^2 \theta_1\Big)d\theta_1^2 + \Big(F_4 - \epsilon ~F_2
f_2^2 {\rm cos}^2 \theta_2\Big)d\theta_2^2\nonumber \nd and we have
defined $\Delta_i$ in the following way: \bg\label{deltadef}
\Delta_1 ~ = ~ \alpha F_2 F_4 {\rm sin}^2 \theta_2, ~~~ \Delta_2 ~ =
~ \alpha F_2 F_3 {\rm sin}^2 \theta_1 \nd At this stage we can
extract the consequence of the fact that both $f_1$ and $f_2$ are
very large. This fits perfectly well with the mirror metric because
$f_i^2$ as well as $f_1f_2$ come with the coefficient $\epsilon$ allowing us to satisfy both SYZ and
susy in the mirror.
This means that if we impose the following constraint:
\bg\label{consf1f2} f_1 f_2 \epsilon ~ \equiv ~ -\alpha \nd i.e both
$f_i$ proportional to $\epsilon^{-1/2}$, it will bring the
cross-terms in the metric to the following suggestive form:
\bg\label{suggest} 2\alpha j_{\phi_1\phi_2} \Big[d\theta_1 d\theta_2
- \Big(d\phi_1 - b_{\phi_1\theta_1} d\theta_1-b_{\phi_1
r}dr\Big)\Big(d\phi_2 - b_{\phi_2\theta_2} d\theta_2-b_{\phi_2
r}dr\Big)\Big] \nd In the limit $b_{\phi_1\alpha} =
b_{\phi_2\alpha} =0$ with $\alpha = r, \theta_i$, \eqref{suggest} is in fact a term of the
deformed conifold! The above conclusion seems rather encouraging,
provided of course \eqref{consf1f2} is satisfied. In the local
limit, similar condition also arose (see the first reference of
\cite{gtpapers}) and we argued therein that as long as we can define
\bg\label{fiear} f_i ~\propto ~ {(-1)^i \langle\alpha\rangle_i \over
\sqrt{\epsilon}} \nd where $\langle\alpha\rangle_i$ depend only on
$\theta_i$ the constraint \eqref{consf1f2} is satisfied. Therefore a
condition like \eqref{consf1f2} works perfectly well in the local
case. Question is, can we satisfy \eqref{consf1f2} also for the
global case?

The answer is now tricky. We demanded that $f_i = f_i(\theta_i)$, otherwise global coordinate transformation like
\eqref{tranbg} {\it cannot} be defined. This means that $F_i$ appearing in the definition of $\alpha$ in \eqref{consf1f2}
will have to be highly constrained. Generically this is not possible\footnote{For the local case $\alpha$ was
defined at $r = r_0$ so this subtlety did not arise and, as we discussed above, we used $\langle\alpha\rangle_i$ to
define $f_i$ so things were perfectly consistent there.} because of the underlying type IIA supersymmetry,
as constraints on the
torsion classes \cite{torsion} will tend to fix $f_i(\theta_i)$,
but in special case this may happen.

The special case arises if we allow $F_2$ to depend on the angular coordinates $\theta_i$
also in such a way that the susy constraints on the torsion classes are still satisfied with $F_2$ given by:
\bg\label{f2def}
F_2(r, \theta_1, \theta_2) ~= ~ - {(\beta_1\beta_2)^{-1} + F_3 F_4 {\rm sin}^2\theta_1 {\rm sin}^2\theta_2 \over
F_4 {\rm cos}^2\theta_1 {\rm sin}^2\theta_2 + F_3 {\rm sin}^2\theta_1 {\rm cos}^2\theta_2}
\nd
where $f_i \equiv {{\beta}_i\over \sqrt{\epsilon}}$. This tells us that the radial dependence of $F_2$ is fixed by
$F_3(r)$ and $F_4(r)$, but the angular dependences are pretty much unfixed because $\beta_i(\theta_i)$ are arbitrary
functions of $\theta_i$ respectively\footnote{Interestingly we can make both $\beta_i$ and $\epsilon$ to be
generic functions of the internal coordinates in such a way that $f_i \equiv {{\beta}_i\over \sqrt{\epsilon}}$
remain functions of $\theta_i$ {\it only}. The only bound on $\epsilon$ would be that it never exceeds 1 over any point 
in the internal space.}.
All these of course should get fixed once we impose the susy constraints on the
torsion classes.
However the above relation \eqref{f2def} already looks tight, but lets move
on and see how far we can go with these kind of arguments. Our next
question would therefore be: is there a way to fix the angular dependences also?

To see how to fix the angular dependences, we can go back to the equivalent local limit of
\eqref{suggest} where the particular way of writing the
metric allows us to make a coordinate rotation to bring the term \eqref{suggest} into the more familar deformed
conifold form \cite{minasian}. This, as we know from \cite{minasian, gtpapers}, is only possible iff {\it other}
terms in the metric remain invariant under the coordinate transformation. If this condition is
imposed globally, then it would imply the
following two relations:
\bg\label{2rels}
&& \beta_1 ~ = ~ \pm\sqrt{F_3 - \alpha j_{\phi_2\phi_2}\over F_2 {\rm cos}^2 \theta_1}\nonumber\\
&& \beta_2 ~ = ~ \mp \sqrt{F_4 - \alpha j_{\phi_1\phi_1}\over F_2
{\rm cos}^2 \theta_1} \nd In the local case, studied in the first
reference of \cite{gtpapers}, relations like \eqref{2rels} are
consistent in the sense that \eqref{consf1f2} is satisfied.
Unfortunately, this is no longer true for the global case
generically because the above relation along with \eqref{consf1f2}
would lead to inconsistent set of equations, and would probably break susy.
Therefore in general
the mirror metric will take the following form: \bg\label{mirmet}
&&ds^2_6 = F_1 dr^2 + {\alpha F_2 \over \Delta_1 \Delta_2}
\Big[d\psi-b_{\psi r}dr + \Delta_1 {\rm cos}~\theta_1 \Big(d\phi_1 -
b_{\phi_1\theta_1} d\theta_1-b_{\phi_1 r}dr\Big)\nonumber\\
&& \quad\quad\quad\quad\quad\quad\quad\quad\quad\quad+ \Delta_2 {\rm
cos}~\theta_2 \Big(d\phi_2 - b_{\phi_2\theta_2} d\theta_2-b_{\phi_2
r}dr\Big)\Big]^2\nonumber\\
&& ~~~~~~ + \alpha j_{\phi_2\phi_2}\Big(d\phi_1 - b_{\phi_1\theta_1}
d\theta_1-b_{\phi_1 r}dr\Big)^2 + \alpha
j_{\phi_1\phi_1}\Big(d\phi_2 - b_{\phi_2\theta_2}
d\theta_2-b_{\phi_2
r}dr\Big)^2\nonumber\\
&& ~~~~~~ -2\alpha j_{\phi_1\phi_2} \Big(d\phi_1 -
b_{\phi_1\theta_1} d\theta_1-b_{\phi_1 r}dr\Big)\Big(d\phi_2 -
b_{\phi_2\theta_2} d\theta_2-b_{\phi_2 r}dr\Big)\nonumber\\ &&
~~~~~~ -2 j_{\phi_1\phi_2} \beta_1 \beta_2~ d\theta_1 d\theta_2 +
\Big(F_3 -  F_2 \beta_1^2 {\rm cos}^2 \theta_1\Big)d\theta_1^2 +
\Big(F_4 - F_2 \beta_2^2 {\rm cos}^2
\theta_2\Big)d\theta_2^2\nonumber\\ \nd Only in very special cases,
where \eqref{2rels} and \eqref{consf1f2} are both simultaneously
satisfied, we expect the mirror to take the following symmetric
form: \bg\label{mirmetspecial} ds^2_6 = && F_1 dr^2 + {\alpha F_2
\over \Delta_1 \Delta_2} \Big[d\psi -b_{\psi r}dr+ \Delta_1 {\rm
cos}~\theta_1 \Big(d\phi_1 - b_{\phi_1\theta_1} d\theta_1-b_{\phi_1
r}dr\Big)
\nonumber\\
&&\quad\quad\quad\quad\quad\quad\quad\quad\;+ \Delta_2 {\rm
cos}~\theta_2 \Big(d\phi_2 - b_{\phi_2\theta_2} d\theta_2-b_{\phi_2
r}dr\Big)\Big]^2\nonumber\\
&& + \alpha j_{\phi_2\phi_2}\Big[d\theta_1^2 + \Big(d\phi_1 -
b_{\phi_1\theta_1} d\theta_1-b_{\phi_1 r}dr\Big)^2\Big] \nonumber\\
&&+ \alpha j_{\phi_1\phi_1}\Big[d\theta_2^2 + \Big(d\phi_2 -
b_{\phi_2\theta_2} d\theta_2-b_{\phi_2
r}dr\Big)^2\Big]\nonumber\\
&& + 2\alpha j_{\phi_1\phi_2}\Big[d\theta_1 d\theta_2 - \Big(d\phi_1
- b_{\phi_1\theta_1} d\theta_1-b_{\phi_1 r}dr\Big)\Big(d\phi_2 -
b_{\phi_2\theta_2} d\theta_2-b_{\phi_2 r}dr\Big)\Big]\nonumber\\ \nd
which is strongly reminiscent of the deformed conifold!
Observe that both the forms of the metrics are finite and well defined.
This tells us that our procedure of making the base large before performing
SYZ \cite{syz} is logical and correct.

On the other hand, the cross term that we developed in the metric appears as
the $B_{\rm NS}$ field in type IIA theory. Expectedly, this B-field is large
and is given by the following form:
\begin{eqnarray}\label{BinIIA}
\widetilde{B}&&=~\alpha
f_1F_3\sin^2\theta_1\left(F_2\cos^2\theta_2+F_4\sin^2\theta_2\right)d\theta_1\wedge d\phi_1\nonumber\\
&&~+\alpha
f_2F_4\sin^2\theta_2\left(F_2\cos^2\theta_1+F_3\sin^2\theta_1\right)d\theta_2\wedge d\phi_2\nonumber\\
&&~+ \left(1-{\epsilon \over \alpha
F_2F_4\sin^2\theta_1\sin^2\theta_2}\right)
\left(f_1\cos\theta_1d\theta_1+f_2\cos\theta_2d\theta_2\right)\wedge d\psi
\end{eqnarray}
In the limit $\epsilon \to 0$, the last two terms are pure gauge. For finite, but small, $\epsilon < 1$ they cannot be
gauged away.
In the local limit (see the first paper of
\cite{gtpapers}) all the $F_i$ were constants, and so $\widetilde{B}$ became a pure gauge when written in terms
of $\langle\alpha\rangle_i$. This doesn't seem to be the case globally, unless of course $F_i$'s are of some specific
forms.

The wrapped D6 brane two-form charges now come partly from the type IIB three-forms and partly from the five-forms.
The three-forms contributions to the IIA two-forms are given by the following
components:
\begin{eqnarray}\label{2form1}
&&\widetilde{F}_{\psi\theta_1}=F_{\phi_1\phi_2\theta_1},\quad
\widetilde{F}_{\psi\theta_2}=F_{\phi_1\phi_2\theta_2},\quad
\widetilde{F}_{\psi r}=F_{\phi_1\phi_2 r},\nonumber\\
&&\widetilde{F}_{\phi_1r}=F_{r\phi_2\psi}+\frac{2j_{\phi_1\phi_2}}{j_{\phi_1\phi_1}}F_{\phi_1
r \psi}+\frac{2j_{\psi\phi_1}}{j_{\phi_1\phi_1}}F_{
r \phi_1\phi_2},\nonumber\\
&&\widetilde{F}_{\phi_2 r}=F_{\phi_1 r\psi}+2\alpha
(j_{\phi_2\psi}j_{\phi_1\phi_1}-j_{\phi_1\phi_2}j_{\phi_1\psi})F_{r\phi_1\phi_2},\nonumber\\
&&\widetilde{F}_{\phi_1\theta_2}=F_{\psi\theta_2\phi_2}+2
\frac{j_{\phi_1\psi}}{j_{\phi_1\phi_1}}F_{\phi_1\phi_2\theta_2},\nonumber\\
&&\widetilde{F}_{\phi_1\theta_1}=2\frac{j_{\phi_1\phi_2}}{j_{\phi_1\phi_1}}F_{\phi_1\theta_1\psi}
+2\frac{j_{\psi\phi_1}}{j_{\phi_1\phi_1}}F_{\phi_1\phi_2\theta_1},\nonumber\\
&&\widetilde{F}_{\phi_2\theta_1}=F_{\psi\phi_1\theta_1}+2\alpha
(j_{\phi_2\psi}j_{\phi_1\phi_1}-j_{\phi_1\phi_2}j_{\phi_1\psi})F_{\phi_1\phi_2\theta_1},\nonumber\\
&&\widetilde{F}_{\phi_2\theta_2}=2\alpha
(j_{\phi_2\psi}j_{\phi_1\phi_1}-j_{\phi_1\phi_2}j_{\phi_1\psi})F_{\phi_1\phi_2\theta_2}
\end{eqnarray}
Similarly, the five-forms contributions to the type IIA two-forms are given in terms of the following components:
\begin{eqnarray}\label{2form2}
&&\widetilde{F}_{\theta_1\theta_2}=F_{\psi\theta_1\theta_2\phi_1\phi_2}+b_{\theta_2\phi_2}F_{\psi\theta_1\phi_1}
+b_{\theta_1\phi_1}F_{\psi\theta_2\phi_2}\nonumber\\
&&\widetilde{F}_{r\theta_1}=F_{r\theta_1\phi_1\phi_2\psi}+b_{\theta_1\phi_1}F_{r\psi\phi_2}
+\left(b_{\phi_2r}+\frac{j_{\phi_1\phi_2}}{j_{\phi_1\phi_1}}b_{r\phi_1}\right)F_{\psi\theta_1\phi_1}
+\frac{j_{\psi\phi_1}}{j_{\phi_1\phi_1}}b_{\theta_1\phi_1}F_{r\phi_1\phi_2}\nonumber\\
&&~~~~~~~~~~~+\left(b_{r\psi}-\frac{j_{\psi\phi_1}}{j_{\phi_1\phi_1}}b_{r\phi_1}\right)F_{\theta_1\phi_1\phi_2}
+\frac{j_{\phi_1\phi_2}}{j_{\phi_1\phi_1}}b_{\theta_1\phi_1}F_{r\psi\phi_1}\nonumber\\
&&\widetilde{F}_{r\theta_2}=F_{r\theta_2\phi_1\phi_2\psi}+b_{r\phi_1}F_{\psi\theta_2\phi_2}
+b_{r\psi}F_{\theta_2\phi_1\phi_2}-b_{\theta_2\phi_2}F_{r\psi\phi_1}
\end{eqnarray}
All the above components are finite and give rise to the required D6-branes charges. However since the B-field is
large, to compensate this in the EOMs we need large G-fluxes in type IIA. These fluxes come exactly from the
extra three- and five-form components \eqref{extra3} and \eqref{extra5} respectively.
These three- and five-form components give rise to twelve components of the four-form
fluxes in IIA namely:
\bg\label{twelve}
&&\widetilde{F}_{r\psi\theta_1\theta_2},~~~
\widetilde{F}_{r\psi\theta_1\phi_1}, ~~~
\widetilde{F}_{r\psi\theta_1\phi_2}, ~~~
\widetilde{F}_{r\psi\theta_2\phi_1}\nonumber\\
&&\widetilde{F}_{r\psi\theta_2\phi_2},~~~
\widetilde{F}_{r\theta_1\theta_2\phi_1}, ~~~
\widetilde{F}_{r\theta_1\theta_2\phi_2}, ~~~
\widetilde{F}_{r\theta_1\phi_1\phi_2}\nonumber\\
&&\widetilde{F}_{r\theta_2\phi_1\phi_2}, ~~~
\widetilde{F}_{\psi\theta_1\theta_2\phi_1},~~~
\widetilde{F}_{\psi\theta_1\theta_2\phi_2},~~~
\widetilde{F}_{\theta_1\theta_2\phi_1\phi_2}
\nd
These components are listed in {\bf Appendix 2} which the readers may refer to for details. Combined with \eqref{BinIIA},
these fluxes lift to M-theory as G-fluxes with components along the spacetime and the eleventh directions respectively.
Interestingly, both the metric \eqref{mirmet} or \eqref{mirmetspecial} along with the two-form flux components
\eqref{2form1} and \eqref{2form2} lift to a geometrical configuration in M-theory, which we expect to have a
$G_2$ structure. This is of course expected because both the non-K\"ahler deformed conifold as well as the wrapped
D6-branes tend to become geometrical configurations when the type IIA coupling is made very 
large\footnote{The fact that we get \eqref{mirmet} instead of \eqref{mirmetspecial} is nothing too surprising. 
One {\it does not} expect to get a K\"ahler deformed conifold in type IIA. This was already clear from the 
pioneering work of \cite{vafa}. Here we not only seem to confirm the statement of \cite{vafa} but also determine 
the precise form of the IIA metric.}.  
In the
following sub-section we will dwell on this in more details.

\subsection{M theory lift, flop transition and type IIA reduction}

The lift of our type IIA mirror configuration to M-theory is rather straighforward. The eleven-directional fibration
is given by gauge fluxes derived from the two-form components \eqref{2form1} and \eqref{2form2}. It is easy to
show that we need only $A_{\phi_i}, A_{\theta_i}$ and $A_r$ components. Using these, the M-theory lift of our IIA
symmetric mirror metric \eqref{mirmetspecial} is:
\begin{eqnarray}\label{mlift}
ds^2_{11}=&&e^{-{2\phi\over 3}}\Bigg\{F_0ds_{0123}^2+F_1 dr^2 + {\alpha F_2
\over \Delta_1 \Delta_2} \Big[d\psi -b_{\psi r}dr\nonumber\\
&&+ \Delta_1 {\rm cos}~\theta_1 \Big(d\phi_1 - b_{\phi_1\theta_1}
d\theta_1-b_{\phi_1 r}dr\Big)+ \Delta_2 {\rm cos}~\theta_2
\Big(d\phi_2 - b_{\phi_2\theta_2} d\theta_2-b_{\phi_2
r}dr\Big)\Big]^2\nonumber\\
&& + \alpha j_{\phi_2\phi_2}\Big[d\theta_1^2 + \Big(d\phi_1 -
b_{\phi_1\theta_1} d\theta_1-b_{\phi_1 r}dr\Big)^2\Big] \nonumber\\
&&+ \alpha j_{\phi_1\phi_1}\Big[d\theta_2^2 + \Big(d\phi_2 -
b_{\phi_2\theta_2} d\theta_2-b_{\phi_2
r}dr\Big)^2\Big]\nonumber\\
&& + 2\alpha j_{\phi_1\phi_2}\Big[d\theta_1 d\theta_2 - \Big(d\phi_1
- b_{\phi_1\theta_1} d\theta_1-b_{\phi_1 r}dr\Big)\Big(d\phi_2 -
b_{\phi_2\theta_2} d\theta_2-b_{\phi_2
r}dr\Big)\Big]\Bigg\}\nonumber\\
&&+e^{4\phi\over 3}\Big[dx_{11}+A_{\phi_1}d\phi_1+A_{\phi_2}d\phi_2+A_{\theta_1}d\theta_1
+A_{\theta_2}d\theta_2+A_rdr\Big]^2
\end{eqnarray}
It is easy to see that the non-symmetric mirror metric \eqref{mirmet} will also lift to M-theory in an
identical way. The local limit of the lift of
\eqref{mirmet} is precisely the one discussed in the first paper of
\cite{gtpapers} and, as discussed therein, we expect the lift of \eqref{mirmet} to have a $G_2$ structure to
preserve supersymmetry. To see this for our case, we have to express the lift of the metric \eqref{mirmet} in terms of
certain one-forms similar to the ones given in \cite{gtpapers} (see also \cite{brandhuber}). Following the
first paper of \cite{gtpapers} we first express the B-fields appearing in the fibration \eqref{mirmet} in terms
of periodic angular coordinates $\lambda_i$ in the following way:
\begin{eqnarray}\label{angles}
&&\tan\lambda_1 ~\equiv ~b_{\phi_1\theta_1},\quad
\tan\lambda_2~\equiv~b_{\phi_2\theta_2},\quad \tan\lambda_3 ~\equiv~b_{\psi
r}\nonumber\\
&&\tan\lambda_4 ~\equiv ~b_{\phi_1 r},\quad \;\,\tan\lambda_5 ~\equiv~ b_{\phi_2 r}
\end{eqnarray}
Using these we can define two sets of one-forms. The first set, called $\sigma_i$ with $i = 1, .., 3$, can be expressed
in terms of $\lambda_i$ as:
\begin{eqnarray}\label{lift1}
\sigma_1&=&\sin\psi_1(d\phi_1-\tan\lambda_4dr)+\sec\lambda_1\cos(\psi_1+\lambda_1)d\theta_1,\nonumber\\
\sigma_2&=&\cos\psi_1(d\phi_1-\tan\lambda_4dr)-\sec\lambda_1\sin(\psi_1+\lambda_1)d\theta_1,\nonumber\\
\sigma_3&=&d\psi_1-\frac{1}{2}\tan\lambda_3dr+\Delta_1\cos\theta_1(d\phi_1-\tan\lambda_1d\theta_1-\tan\lambda_4dr)
\end{eqnarray}
and the second set can be expressed in terms of $\lambda_i$ as:
\begin{eqnarray}\label{lift2}
\Sigma_1&=&-\sin\psi_2(d\phi_2-\tan\lambda_5dr)+\sec\lambda_2\cos(\psi_2+\lambda_2)d\theta_2,\nonumber\\
\Sigma_2&=&-\cos\psi_2(d\phi_2-\tan\lambda_5dr)-\sec\lambda_2\sin(\psi_2+\lambda_2)d\theta_2,\nonumber\\
\Sigma_3&=&d\psi_2+\frac{1}{2}\tan\lambda_3dr-\Delta_2\cos\theta_2(d\phi_2-\tan\lambda_2d\theta_2-\tan\lambda_5dr)
\end{eqnarray}
At this stage one may compare these two sets of one-forms to the ones given by eq. (6.2) and eq. (6.3) in the first
paper of \cite{gtpapers}. The definitions of $\psi_1$ and $\psi_2$ follow exactly as in \cite{gtpapers}, i.e
\bg\label{psi12}
d\psi ~ = ~ d\psi_1 - d\psi_2, ~~~~~~~~~ dx_{11} ~ = ~ d\psi_1 + d\psi_2
\nd
Furthermore we can perform the following rotation of the coordinates:
\begin{equation}\label{rotc}
\begin{pmatrix} {\cal D}\phi_2 \\ d\theta_2\end{pmatrix} ~ \to ~
\begin{pmatrix} \cos~\psi_0 & -\sin~\psi_0 \\ \sin~\psi_0 & ~~\cos~\psi_0 \end{pmatrix}
\begin{pmatrix} {\cal D}\phi_2 \\ d\theta_2\end{pmatrix}
\end{equation}
with ${\cal D}\phi_2 \equiv d\phi_2 - b_{\phi_2\theta_2} d\theta_2-b_{\phi_2r}dr$ and $\psi_0$ a constant. If we make
this transformation to the symmetric mirror metric of type IIA \eqref{mirmetspecial}, this will lift to M-theory not as
\eqref{mlift}, but to a more {\it suggestive} configuration:
\begin{eqnarray}\label{mliftnow}
ds^2_{11}=&&e^{-{2\phi\over 3}}\Bigg\{F_0ds_{0123}^2+F_1 dr^2 + {\alpha F_2
\over \Delta_1 \Delta_2} \Big[d\psi -b_{\psi r}dr - b_{\psi\theta_2} d\theta_2\nonumber\\
&&+ \Delta_1 {\rm cos}~\theta_1 \Big(d\phi_1 - b_{\phi_1\theta_1}
d\theta_1-b_{\phi_1 r}dr\Big)+ \Delta_2 {\rm cos}~\theta_2 {\rm cos}~\psi_0
\Big(d\phi_2 - b_{\phi_2\theta_2} d\theta_2-b_{\phi_2
r}dr\Big)\Big]^2\nonumber\\
&& + \alpha j_{\phi_2\phi_2}\Big[d\theta_1^2 + \Big(d\phi_1 -
b_{\phi_1\theta_1} d\theta_1-b_{\phi_1 r}dr\Big)^2\Big] \nonumber\\
&&+ \alpha j_{\phi_1\phi_1}\Big[d\theta_2^2 + \Big(d\phi_2 -
b_{\phi_2\theta_2} d\theta_2-b_{\phi_2
r}dr\Big)^2\Big]\nonumber\\
&& + 2\alpha j_{\phi_1\phi_2}{\rm cos}~\psi_0\Big[d\theta_1 d\theta_2 - \Big(d\phi_1
- b_{\phi_1\theta_1} d\theta_1-b_{\phi_1 r}dr\Big)\Big(d\phi_2 -
b_{\phi_2\theta_2} d\theta_2-b_{\phi_2
r}dr\Big)\Big]\nonumber\\
&& + 2\alpha j_{\phi_1\phi_2}{\rm sin}~\psi_0\Big[\Big(d\phi_1- b_{\phi_1\theta_1}
d\theta_1-b_{\phi_1 r}dr\Big) d\theta_2
+ \Big(d\phi_2 - b_{\phi_2\theta_2} d\theta_2-b_{\phi_2r}dr\Big)d\theta_1\Big]\Bigg\}\nonumber\\
&&+e^{4\phi\over 3}\Big[dx_{11}+{\widetilde A}_{\phi_1}d\phi_1+{\widetilde A}_{\phi_2}d\phi_2+
{\widetilde A}_{\theta_1}d\theta_1
+{\widetilde A}_{\theta_2}d\theta_2+{\widetilde A}_rdr\Big]^2
\end{eqnarray}
where we have introduced a $B$-field fibration using
$b_{\psi\theta_2} \equiv \Delta_2 ~{\rm sin}~\psi_0 {\rm cos}~\theta_2$ to modify the $d\psi$ fibration structure.
The eleven-dimensional fibration structure will also change accordingly because we can always express the
$A_{\phi_2} d\phi_2$ term in $dx_{11}$ of \eqref{mlift} using ${\cal D}\phi_2$. Thus the overall eleven-dimensional
fibration will retain its form but with shifted $A_\mu$ fields denoted above by the ${\widetilde A}_\mu$ fields.
In terms of the fibration components of \eqref{mlift} one can
show that ${\widetilde A}_{\phi_1} = A_{\phi_1}, {\widetilde A}_{\theta_1} = A_{\theta_1}$ and the rest of the
components can be presented in the following matrix form:
\begin{equation}\label{matricre}
\begin{pmatrix}{\widetilde A}_{\phi_2}\\ {}&{}&{} \\ {\widetilde A}_{\theta_2}\\ {}&{}&{}\\
{\widetilde A}_{r} \end{pmatrix} ~ = ~
\begin{pmatrix} \cos\psi_0 + b_{\phi_2\theta_2}\sin\psi_0 & \sin\psi_0 & 0\\ {}&{}&{}\\
-(1+b^2_{\phi_2\theta_2})\sin\psi_0 & \cos\psi_0 - b_{\phi_2\theta_2}\sin\psi_0 & 0\\ {}&{}&{}\\
b_{\phi_2 r}(1-
\cos\psi_0 - b_{\phi_2\theta_2}\sin\psi_0) & -b_{\phi_2 r} \sin\psi_0 &1 \end{pmatrix}
\begin{pmatrix}{A}_{\phi_2}\\ {}&{}&{}\\ {A}_{\theta_2}\\ {}&{}&{}\\ {A}_{r} \end{pmatrix}
\end{equation}
Additionally with the above modification, the
above metric is surprisingly close to the uplift of a non-K\"ahler deformed conifold metric with wrapped D6-branes
to M-theory provided we can make an additional substitution\footnote{This is compatible with the
underlying $G_2$ structure. An analysis of the $G_2$ torsion classes, along the lines of
the first paper in \cite{gtpapers}, will reveal this. We will discuss this more soon.} in \eqref{mliftnow}:
\bg\label{psis}
\psi_0 ~ ~\to ~~ \psi
\nd
Making such a substitution may lead one to think that the $\psi$ isometry that we have in \eqref{mlift} is removed.
This is {\it not} the case locally
with the non-K\"ahler deformed conifold because the extra B-field component $b_{\psi\theta_2}$
in the $d\psi$ fibration structure of \eqref{mliftnow} as well as the vector fields ${\widetilde A}_\mu$
in the $dx_{11}$ fibration structure transform non-trivially under
shift in $\psi$ to restore the isometry. One may also do a somewhat similar rotation like \eqref{rotc} to the
non-symmetric type IIA metric \eqref{mirmet} and bring it in a more suggestive format.

The rotation \eqref{rotc} should now be captured by
the one-forms \eqref{lift1} and \eqref{lift2} appropriately. In fact only the
second set of one-forms \eqref{lift2} is related to the
change \eqref{rotc}.
%These changes can be easily worked out and, to avoid
%cluttering of formulae, we will rename the changed one-forms \eqref{lift2} as $\Sigma_i$ also.
Thus the one-forms for
our purposes will be ($\sigma_i, \Sigma_i$) with $\Sigma_i$ to be viewed as the one got from \eqref{rotc} directly.
Using these one-forms
we can rewrite the M-theory metric in two possible ways. The first one is the lift of the non-symmetric type IIA
metric \eqref{mirmet} under the rotation \eqref{rotc} and transformation \eqref{psis}:
\begin{eqnarray}\label{liftnow}
ds_7^2 && = g_rdr^2+g_1(\sigma_3+\Sigma_3)^2+g_2(\sigma_3-\Sigma_3)^2\nonumber\\
&&+~g_3(\sin\psi_1\sigma_1+\cos\psi_1\sigma_2)^2
+\widetilde{g}_3(\cos\psi_1\sigma_1-\sin\psi_1\sigma_2)^2\nonumber\\
&&+~g_4(\sin\psi_2\Sigma_1+\cos\psi_2\Sigma_2)^2
+\widetilde{g}_4(\cos\psi_2\Sigma_1-\sin\psi_2\Sigma_2)^2\nonumber\\
&&+~g_5(\sin\psi_1\sigma_1+\cos\psi_1\sigma_2)(\sin\psi_2\Sigma_1+\cos\psi_2\Sigma_2)\nonumber\\
&&-~\widetilde{g}_5(\cos\psi\sigma_1-\sin\psi_1\sigma_2)(\cos\psi_2\Sigma_1-\sin\psi_2\Sigma_2)
\end{eqnarray}
where we have defined the coefficients $g_i, {\widetilde g}_i$ as:
\begin{eqnarray}\label{ggdef}
&&g_r=e^{-2\phi/3}F_1,\quad g_1=e^{4\phi/3},\quad
g_2=e^{-2\phi/3}\frac{\alpha F_2}{\Delta_1\Delta_2},\quad g_3=\alpha
j_{\phi_2\phi_2},
\nonumber\\
&&\widetilde{g}_3=F_3-F_2\beta_1^2\cos^2\theta_1,\quad g_4=\alpha
j_{\phi_1\phi_1},\quad
\widetilde{g}_4=F_4-F_2\beta_2^2\cos^2\theta_2\nonumber\\
&&g_5=2\alpha j_{\phi_1\phi_2},\quad
\widetilde{g}_5=2\beta_1\beta_2j_{\phi_1\phi_2}
\end{eqnarray}
The second way to rewrite the metric is a little more suggestive of the way to perform the flop operation on the M-theory
manifold and has a nice form for the symmetric case \eqref{mirmetspecial}, again under \eqref{rotc} and \eqref{psis}.
The local form of this has already appeared in the first reference of \cite{gtpapers}, and the readers may
want to look at that for more details. In fact we can rewrite \eqref{liftnow} in the following
way also.
Here we will simply quote the generic ansatze using parameters $\alpha_i$ and $\zeta$:
\bg\label{sugmet}
ds_7^2 = \alpha_1^2 \sum_{a=1}^2 (\sigma_a + \zeta \Sigma_a)^2 + \alpha_2^2 \sum_{a=1}^2 (\sigma_a - \zeta \Sigma_a)^2
+ \alpha_3^2 (\sigma_3 + \Sigma_3)^2 + \alpha_4^2 (\sigma_3 - \Sigma_3)^2 + \alpha_5^2 dr^2\nonumber\\
\nd
The above is a familiar form by which any $G_2$ structure metric could be expressed. Once we switch off $\lambda_i$ the
manifolds has a $G_2$ holonomy. The coefficients $\alpha_i$ and $\zeta$ are not arbitrary. They are fixed by the EOM
and, for the case \eqref{mirmetspecial}, they take the following values:
\bg\label{alphazeta}
&& \alpha_1 = {1\over 2} e^{-{\phi\over 3}}\sqrt{2\alpha\left(j_{\phi_2\phi_2} + {j_{\phi_1\phi_2}\over \zeta}\right)},
~~~~~ \alpha_2 = {1\over 2} e^{-{\phi\over 3}}\sqrt{2\alpha\left(j_{\phi_2\phi_2}
- {j_{\phi_1\phi_2}\over \zeta}\right)}\nonumber\\
&& \alpha_3 = e^{2\phi\over 3}, ~~~~ \alpha_4 = e^{-{\phi\over 3}} \sqrt{\alpha F_2 \over \Delta_1 \Delta_2}, ~~~~
\alpha_5 =  e^{-{\phi\over 3}} \sqrt{F_1}, ~~~~ \zeta = \sqrt{j_{\phi_1\phi_1}\over j_{\phi_2\phi_2}}
\nd
Before proceeding further let us pause for a while and ask whether the above set of manipulations would preserve
supersymmetry. From type IIA point of view we have done the following:

\vskip.1in

\noindent $\bullet$ Shift of the coordinates ($\psi, \phi_i$) using variables $f_i(\theta_i)$. This shifting of the
coordinates mixes non-trivially all the three isometry directions as described in \eqref{tranbg}.

\noindent $\bullet$ Shift the metric along $\psi$ direction by the variable $\epsilon$, as given in the second line
of \eqref{metcom}. This variable doesn't have to be too small in the global limit. As long as it is smaller than 1 
it'll suffice.

\noindent $\bullet$ Make SYZ transformations along the new shifted directions. Thus the three T-dualities are
{\it not} made along the three original isometry directions.

\noindent $\bullet$ In the new metric of IIA make a further rotation along the ($\theta_2, \phi_2$)
directions using a $2\times 2$
matrix given as \eqref{rotc}. The matrix is described using a constant angular variable $\psi_0$.

\noindent $\bullet$ Finally in the transformed metric convert $\psi_0$ to $\psi$ as in \eqref{psis}.

\vskip.1in

\noindent However due to steps 2, 4 and 5 above, it is not guaranteed that the metric will preserve supersymmetry.
Furthermore one might also question whether the SYZ operation itself could preserve supersymmetry. Therefore to verify
this we have evaluated all the torsion classes for this background in sec 3.2 of \cite{chen2}. The supersymmetry
constraints are given by eq. (3.21) and Appendix B of \cite{chen2}. These set of equations along with the constraint
equation (3.15) of \cite{chen2}  are enough to guarantee supersymmetry in the type IIA (or the equivalent
M-theory) background.

Once the issue of supersymmetry is resolved, we go to the flop operation.
The operation of flop on the above metric \eqref{sugmet} has already been discussed in details in sec. 7 of the
first reference of \cite{gtpapers}. Using similar techniques for \eqref{liftnow},
after the flop we expect the metric to look like:
\begin{eqnarray}\label{aflop}
ds_7^2=a_1(\sigma_1^2+\sigma_2^2)+a_2(\Sigma_1^2+\Sigma_2^2)+a_3(\sigma_3+\Sigma_3)^2+a_4(\sigma_3-\Sigma_3)^2+a_5dr^2
\end{eqnarray}
with $a_i$, $i=1, ..., 5$ are some coefficients to be determined. Due to the global nature of our metric,
the operation of flop can be performed by
a class of transformations parametrized by the values of
$a$, $b$ etc. in the following way:
\begin{eqnarray}\label{lbaaz}
&&\sigma_1\mapsto a\sigma_1+b\Sigma_1,\quad \Sigma_1\mapsto
e\sigma_1+f\Sigma_1,\nonumber\\
&&\sigma_2\mapsto c\sigma_2+d\Sigma_2,\quad \Sigma_2\mapsto
g\sigma_2+h\Sigma_2,\nonumber\\
&&\sigma_3+\Sigma_3\mapsto \sigma_3-\Sigma_3, \quad
\sigma_3-\Sigma_3\mapsto \sigma_3+\Sigma_3
\end{eqnarray}
Now comparing \eqref{sugmet} and \eqref{liftnow} one can pretty much fix the coefficients $c, d$ etc. in terms of
$a, b$ in the following way:
\bg\label{cdef}
&& c=a\sqrt{\frac{k^2G_2+kG_3+G_1}{\omega^2G_5+\omega
G_6+G_4}},~~~~~~~ d=b\sqrt{\frac{\mu^2G_2+\mu G_3+G_1}{\tau^2G_5+\tau
G_6+G_4}}\nonumber\\
&& e=ak,~~~~ g=c\omega,~~~~ f=b\mu,~~~~ h=d\tau
\nd
where $a, b$ can in turn be fixed by computing the $G_2$-torsion classes and demanding supersymmetry (see also
\cite{chen2}).
The other coefficients appearing above, namely, $k$,
$\omega$, $\mu$, $\tau$ satisfy the following equations:
\begin{eqnarray}
&&2G_1+2k\mu G_2+(k+\mu)G_3=0, \quad~ G_7+k\tau G_8+\tau G_9+kG_{10}=0,\nonumber\\
&&2G_4+2\tau \omega G_5+(\tau+\omega)G_6=0, \quad G_7+\omega\mu
G_8+\omega G_9+\mu G_{10}=0.
\end{eqnarray}
whose solutions are fixed by the following values of $G_i$ determined from the $G_2$ structure metric \eqref{liftnow}
using ($g_i, \widetilde{g}_i$) defined earlier in \eqref{ggdef}:
\begin{eqnarray}
&&G_1=g_3\sin^2\psi_1+\widetilde{g}_3\cos^2\psi_1,\quad\quad
G_2=g_4\sin^2\psi_2+\widetilde{g}_4\cos^2\psi_2, \nonumber\\
&&G_3=g_5\sin\psi_1\sin\psi_2-\widetilde{g}_5\cos\psi_1\cos\psi_2,\nonumber\\
&&G_4=g_3\cos^2\psi_1+\widetilde{g}_3\sin^2\psi_1,\quad\quad
G_5=g_4\cos^2\psi_2+\widetilde{g}_4\sin^2\psi_2, \nonumber\\
&&G_6=g_5\cos\psi_1\cos\psi_2-\widetilde{g}_5\sin\psi_1\sin\psi_2,\nonumber\\
&&G_7=(g_3-\widetilde{g}_3)\sin\psi_1\cos\psi_1,\quad\;
G_9=g_5\sin\psi_1\cos\psi_2+\widetilde{g}_5\cos\psi_1\sin\psi_2,\nonumber\\
&&G_8=(g_4-\widetilde{g}_4)\sin\psi_2\cos\psi_2,\quad\;
G_{10}=g_5\cos\psi_1\sin\psi_2+\widetilde{g}_5\sin\psi_1\cos\psi_2.\nonumber\\
\end{eqnarray}
Using all the above relations, the $a_i$ coefficients in the M-theory metric after flop transition \eqref{lbaaz} can
be determined in terms of ($a, b$). The final form of the metric therefore is given by:
\begin{eqnarray}\label{flopmet}
ds_{11}^2&&= e^{-{2\phi\over 3}}F_0 ds_{0123}^2+g_rdr^2+g_1(\sigma_3-\Sigma_3)^2+g_2(\sigma_3+\Sigma_3)^2\nonumber\\
&&+a^2(k^2G_2+kG_3+G_1)(\sigma_1^2+\sigma_2^2)+b^2(\mu^2G_2+\mu
G_3+G_1)(\Sigma_1^2+\Sigma_2^2)\nonumber\\
\end{eqnarray}
We are now one step away from getting the type IIA metric from the above metric.
Reducing along $x_{11}$ the metric takes the following form in type IIA theory:
\begin{eqnarray}\label{IIAmetric}
ds_{10}^2&& =F_0ds_{0123}^2+F_1dr^2+e^{2\phi}\Big[d\psi-b_{\psi
\mu}dx^\mu+\Delta_1 {\rm cos}~\theta_1 \Big(d\phi_1 - b_{\phi_1\theta_1}
d\theta_1-b_{\phi_1 r}dr\Big)\nonumber\\
&&\quad\quad\quad\quad\quad\quad\quad\quad\quad\quad +
\widetilde{\Delta}_2 {\rm cos}~\theta_2 \Big(d\phi_2 - b_{\phi_2\theta_2}
d\theta_2-b_{\phi_2 r}dr\Big)\Big]^2\nonumber\\
&&~+e^{2\phi\over 3}a^2(k^2G_2+kG_3+G_1)\Big[d\theta_1^2+(d\phi_1^2-b_{\phi_1\theta_1}d\theta_1-b_{\phi_1
r}dr)^2\Big]\nonumber\\
&&~+e^{2\phi\over 3}b^2(\mu^2G_2+\mu
G_3+G_1)\Big[d\theta_2^2+(d\phi_2^2-b_{\phi_2\theta_2}d\theta_2-b_{\phi_2
r}dr)^2\Big]
\end{eqnarray}
which has an amazing similarity with the warped resolved conifold! The above metric is completely global and
supersymmetric\footnote{Of course there is a further UV completion that we don't discuss here. The
UV completion should follow in the same vein as studied recently for the
Klebanov-Strassler case in \cite{FEP, jpsi}.}. As before, the full torsion class analysis for this is
given in eq. (3.25) of \cite{chen2}. Together with equations (3.15), (3.21) and (3.25) of \cite{chen2} we can
pretty much get most of the warp factor components (plus the other parameters) to demand supersymmetry for all the
above backgrounds. Any remaining set of unconstrained parameters would allow us to get a class of gauge
theory deformations that span the landscape of solutions in the geometric transition set-up. We will discuss more
on this landscape soon (see also figure 3 of \cite{chen2}).

Therefore the type IIA background
should be viewed as the gravity dual in the IR for the gauge theory on wrapped D6-branes before
geometric transition. In this background there are no six-branes. The wrapped D6-branes have {\it dissolved} in the
geometry, and is replaced by the following one-form flux components:
\bg\label{1ffc}
 A=\Delta_1 {\rm cos}~\theta_1 \Big(d\phi_1 -
b_{\phi_1\theta_1} d\theta_1-b_{\phi_1 r}dr\Big)- \widetilde{\Delta}_2 {\rm
cos}~\theta_2 \Big(d\phi_2 - b_{\phi_2\theta_2} d\theta_2-b_{\phi_2
r}dr\Big)
\nd
with $\widetilde{\Delta}_2$ is a slight deformation of $\Delta_2$ appearing from the rotation \eqref{rotc}
before the flop operation. The type IIA background also supports an effective dilaton, that measures the
IIA coupling, and is given by:
\bg\label{edil}
\phi_{\rm eff}~ = ~ \frac{3}{4}\ln(g_2)
\nd
Before we end this section, there are a few loose ends that need to be tied up. The first one is related to the M-theory
G-fluxes. These G-fluxes stem from \eqref{twelve} and \eqref{BinIIA} in type IIA,
and they are in general large\footnote{As we discussed before $\epsilon$ in \eqref{metcom} or \eqref{consf1f2} is a small
but {\it finite} number less than 1 (otherwise the signature will change), 
the type IIA flux components \eqref{BinIIA} and \eqref{twelve} will be large but finite.
Choosing a particular 
value of $\epsilon$ and then demanding supersymmetry will consequently fix the resulting fluxes. A range of choices for 
$\epsilon$ will give a class of backgrounds which are dual to certain continuous deformations parametrised 
by $\epsilon$ in the gauge theory side. We will discuss more about the class of backgrounds later.}.
In the
local picture both \eqref{BinIIA} as well as \eqref{twelve} components were all pure gauges, and therefore they
did not contribute to the background. Here we expect they would, and therefore we need to see how these fluxes behave
under:

\vskip.1in

\noindent $\bullet$ The rotation of coordinates \eqref{rotc} with shift \eqref{psis}, and

\noindent $\bullet$ The flop transformation \eqref{lbaaz}.

\vskip.1in

\noindent Both these effects can be worked out if we can express our fluxes \eqref{twelve} and \eqref{BinIIA}
completely in terms of the one-forms \eqref{lift1} and \eqref{lift2}. As we noted before, the rotation
\eqref{rotc} and shift \eqref{psis} is appropriately captured by
the one-forms \eqref{lift2}. Therefore to compensate both the changes, namely rotation
\eqref{rotc} (with shift \eqref{psis})
and the flop \eqref{lbaaz}, all we need is to express the M-theory lift of the fluxes in terms of
\eqref{lift1} and \eqref{lift2}. Any {\it additional} $\psi$ dependent contributions will come out from
susy requirement.
%The latter can be easily performed by first expressing the G-fluxes
%in terms of \eqref{lift2} and then change $\Sigma_i$ to the transformed $\Sigma_i$ (recall that we are using the same
%notation for $\Sigma_i$ and its transformed version).

To achieve all this, we can express the differential coordinates completely in terms of $\sigma_i$ and $\Sigma_i$.
Since there are seven differential coordinates ($dr, d\theta_1, d\phi_1, d\theta_2, d\phi_2, d\psi_1, d\psi_2$) but
six one-forms ($\sigma_i, \Sigma_i$), we can assume $dr$ goes to itself, and then the rest of the differential
forms map to the ($\sigma_i, \Sigma_i$) in the following way:
\begin{eqnarray}
&&d\theta_1=\cos\psi_1\sigma_1-\sin\psi_1\sigma_2,\quad
d\theta_2=\cos\psi_2\Sigma_1-\sin\psi_2\Sigma_2,\nonumber\\
&&d\phi_1-\tan\lambda_4dr=\sec\lambda_1\Big[\sin(\psi_1+\lambda_1)\sigma_1+\cos(\psi_1+\lambda_1)
\sigma_2\Big],\nonumber\\
&&d\phi_2-\tan\lambda_5dr=-\sec\lambda_2\Big[\sin(\psi_2-\lambda_2)\Sigma_1+\cos(\psi_2-\lambda_2)
\Sigma_2\Big],\nonumber\\
&&d\psi-\tan\lambda_3dr=\sigma_3-\Sigma_3-\Delta_1 {\rm
cos}~\theta_1 \Big(\sin\psi_1\sigma_1+\cos\psi_1\sigma_2\Big)\nonumber\\
&&\quad\quad\quad\quad\quad\quad\quad\quad\quad\quad\quad\;+
\Delta_2 {\rm cos}~\theta_2
\Big(\sin\psi_2\Sigma_1+\cos\psi_2\Sigma_2\Big)\nonumber\\
&&dx_{11}=\sigma_3+\Sigma_3+\Delta_1 {\rm
cos}~\theta_1 \Big(\sin\psi_1\sigma_1+\cos\psi_1\sigma_2\Big)\nonumber\\
&&\quad\quad\quad\quad\quad\quad\quad- \Delta_2 {\rm cos}~\theta_2
\Big(\sin\psi_2\Sigma_1+\cos\psi_2\Sigma_2\Big)
\end{eqnarray}
Under the rotation, shift and flop the flux components mix in rather
non-trivial way.
We therefore expect, after the IIA reduction from M-theory, the three-form and four-form flux components of type
IIA \eqref{BinIIA} and \eqref{twelve} respectively before geometric transition go to {\it new} three- and four-form
flux components. They can be expressed as:
\begin{eqnarray}\label{BIIA2}
{B}_{\rm now}~=~ \bar{b}_{ij}dx^i\wedge dx^j, ~~~~~~~~
{F}_{\rm now} ~= ~ \bar{f}_{ijkl}dx^i\wedge dx^j\wedge dx^k\wedge dx^l
\end{eqnarray}
where $x^{i,j,k,l}=r, \theta_i, \phi_i, \psi$ and
$\bar{b}_{ij}$ and $\bar{f}_{ijkl}$
are not only functions of ($r, \theta_i$) but now also of $\psi$ because of the shift \eqref{psis} and rotation
\eqref{rotc}.

%poporam

\subsection{Type IIB after mirror transition}

With the type IIA picture at hand, we are now at the last chain of the duality transformation that will give us the
supergravity dual of the confining gauge theory on the wrapped D5-branes. Our starting points are now:

\vskip.1in

\noindent $\bullet$ The type IIA metric \eqref{IIAmetric}.

\noindent $\bullet$ The remnant of the D6-brane charges, i.e the one-form fluxes \eqref{1ffc}.

\noindent $\bullet$ The type IIA string coupling, or the dilaton \eqref{edil}, and

\noindent $\bullet$ The $B_{\rm NS}$ and the $F_4$ fluxes \eqref{BIIA2}
from the remnant of the IIB shift transformations.

\vskip.1in

\noindent Before moving ahead, let us make two observations. The first is
that now we {\it do} have components like $\bar{b}_{\phi_1\phi_2}, \bar{b}_{\psi\phi_i}$ in \eqref{BIIA2}.
This would mean that the
mirror type IIB should become non-geometric! This is what one would have expected generically, and our analysis
does confirm this\footnote{In fact this tells us that the generic solution spaces we get in type IIB are non-geometric
manifolds. For certain choices of parameters (B-fields, and metric components) we can get geometric manifolds like
Klebanov-Strassler \cite{KS} or Maldacena-Nunez \cite{MN}.
This is almost like the parameter space of \cite{butti} but now much bigger,
and allowing {\it both} geometric and non-geometric manifolds that cover various
branches of the dual gauge theories. One may also note that the susy constraints from the torsion
class analysis in equation (3.15), (3.21) and (3.25) of \cite{chen2} would keep some of the parameters unfixed (one example would be the $\epsilon$ parameter that we discussed earlier). 
Therefore
these parameters could be {\it varied} to get a class of gauge theory deformations. Typically the duals of
these deformations are non-geometric manifolds. This is expressed in figure 3 of \cite{chen2}.}.
Observe that locally, as in \cite{gtpapers}, this aspect of non-geometricity was not visible because
most of the extra fluxes were pure gauges. In the global case, the system is rather non-trivial and the dual gravitational
description may become non-geometric. Question now is whether we can look for a special case where we can study the
system as a {\it geometric} manifold. It turns out at the orientifold point there might be a situation where we can
switch off the extra flux components and consider only the standard B-field components. Recall that due to various
rotations \eqref{rotc} and shifting \eqref{tranbg} and \eqref{psis}
the orientifolding is more involved, as all the internal coordinates
are mixed up in these transformations. However this may not generically remove all the necessary components.
Therefore to simplify the situation, in the following, we will study the
type IIB mirror by first keeping:
\bg\label{nihar} \bar{b}_{\phi_1\phi_2} ~ = ~ \bar{b}_{\psi\phi_i} ~ = ~0\nd
so that the mirror could be geometric. Switching on \eqref{nihar} in the IIA scenario will then make the system
non-geometric.

\noindent Secondly,
note that except for the $B_{\rm NS}$ and the $F_4$ fluxes, rest of the components of the metric or the
one-form fluxes, or even the dilaton are all finite. The $B_{\rm NS}$ and the $F_4$ fluxes are large, and in the limit
$\epsilon$ in \eqref{metcom} is a small but finite integer, these would also be finite (but large). To proceed further
let us define:
\bg\label{defui}
{\cal D}\phi_1 \equiv d\phi_1 - b_{\phi_1\theta_1} d\theta_1-b_{\phi_1 r}dr,~~~~~~
{\cal D}\phi_2 \equiv d\phi_2 - b_{\phi_2\theta_2} d\theta_2-b_{\phi_2 r}dr
\nd
Using this, the type IIA metric can be rewritten as:
\begin{eqnarray}\label{IIAmetre}
ds_{10}^2&& =F_0ds_{0,1,2,3}^2+F_1dr^2+e^{2\phi}\Big(d\psi-b_{\psi
\mu}dx^\mu+\Delta_1 {\rm cos}~\theta_1 ~{\cal D}\phi_1 +
\widetilde{\Delta}_2 {\rm cos}~\theta_2~ {\cal D}\phi_2\Big)^2\nonumber\\
&&~+ {\cal F}_1 \Big(d\theta_1^2+ {\cal D}\phi_1^2\Big) +
{\cal F}_2 \Big(d\theta_2^2+ {\cal D}\phi_2^2\Big)
\end{eqnarray}
In this form the non-K\"ahlerity is obvious in terms of the fibrations ${\cal D}\phi_i$ and
the resolution parameters of the two two-cycles are determined completely in terms of ${\cal F}_i$ as:
\bg\label{calfo}
{\cal F}_1 = e^{2\phi\over 3}a^2(k^2G_2+kG_3+G_1), ~~~~
{\cal F}_2 = e^{2\phi\over 3}b^2(\mu^2G_2+\mu G_3+G_1)
\nd
It is now clear that to determine the type IIB mirror using SYZ \cite{syz} we have to make the base bigger as before.
The manifold \eqref{IIAmetre} still retains isometries along ($\phi_i, \psi$), so after we enlarge the base we can
perform SYZ in the usual way\footnote{This is not as simple as it sounds. As we saw above, due to rotation \eqref{rotc}
and shift \eqref{psis}, the fluxes may develop dependences on $\psi$ coordinate. A way out of this would be to
express the fluxes in terms of an average $\langle\psi\rangle$ and then perform the SYZ transformations. We can then
demand supersymmetry in the
final IIB configuration by analysing the torsion classes, exactly as we did for all the above
cases. This way all intermediate configurations would be supersymmetric.}.
Since these details are rather straightforward to work out, we will not redo them again
now. To put the type IIB metric in some suggestive format, let us define the following quantities:
\begin{eqnarray}\label{newthings}
&&\bar{\alpha}=\Big(\bar{j}_{\phi_1\phi_1}\bar{j}_{\phi_2\phi_2}-\bar{j}_{\phi_1\phi_2}+
\bar{b}_{\phi_1\phi_2}^2\Big)^{-1}, ~~~~~ \widetilde{\cal D}\psi \equiv d\psi+g_{r\psi}dr \nonumber\\
&& \widetilde{\cal D}\phi_1 \equiv \sqrt{g_{\phi_1\phi_1}\over g_{\theta_1\theta_1}}
\Big(d\phi_1+g_{\phi_1\theta_1}d\theta_1+g_{r\phi_1}dr\Big), ~~~
\widetilde{\cal D}\phi_2 \equiv \sqrt{g_{\phi_2\phi_2}\over g_{\theta_2\theta_2}}
\Big(d\phi_2+g_{\phi_2\theta_2}d\theta_2+g_{r\phi_2}dr\Big)\nonumber\\
\end{eqnarray}
where
$\bar{j}_{\mu\nu}$ denote the components of the IIA metric \eqref{IIAmetre}, and $g_{\mu\nu}$ are defined in terms of
$\bar{j}_{\mu\nu}$ in {\bf Appendix 3}. Using these definitions, and taking $b_{r\phi_i} = b_{r\psi} = 0$,
the mirror manifold in type IIB theory
takes the following form:
\begin{eqnarray}\label{iibmetfinal}
ds^2 & = & F_0^2ds_{0,1,2,3}^2+g_{rr}dr^2  +g_{\psi\psi}\Big(\widetilde{\cal D}\psi
+ \widehat{\Delta}_1 ~\widetilde{\cal D}\phi_1
+ \widehat{\Delta}_2 ~\widetilde{\cal D}\phi_2\Big)^2\\
&& +g_{\theta_1\theta_1}\Big(d\theta_1^2 + \widetilde{\cal D}\phi_1^2\Big)
+g_{\theta_2\theta_2}\Big(d\theta_2^2 + \widetilde{\cal D}\phi_2^2\Big)
+g_{\theta_1\theta_2}\Big(d\theta_1d\theta_2 + \widehat{\Delta}_3
~\widetilde{\cal D}\phi_1 \widetilde{\cal D}\phi_2\Big)\nonumber
\nd
which looks surprisingly close to the warped resolved-deformed conifold! Clearly the manifold is non-K\"ahler and
$\hat{\Delta}_i$ are defined as:
\bg\label{hatdeldef}
\widehat{\Delta}_1 \equiv \sqrt{g_{\theta_1\theta_1}g^2_{\psi\phi_1}\over g_{\phi_1\phi_1}}, ~~~~~
\widehat{\Delta}_2 \equiv \sqrt{g_{\theta_2\theta_2}g^2_{\psi\phi_2}\over g_{\phi_2\phi_2}}, ~~~~~
\widehat{\Delta}_3 \equiv \sqrt{g_{\theta_1\theta_1} g_{\theta_2\theta_2}g^2_{\phi_1\phi_2}\over
g_{\phi_1\phi_1}g_{\phi_2\phi_2}g^2_{\theta_1\theta_2}}
\nd
The type IIB fluxes are rather involved, but they could be worked out exactly as in {\bf Appendix 2}. We will
not do so here, but discuss their implications in our follow-up paper \cite{toappear}. It is interesting that the
solutions that we get in type IIA as well as type IIB for the gravity duals look very close to what has been
advocated in the literature so far in the limit where we switch off certain
components of the $\bar{b}$-fields as well as $b_{r\phi_i}, b_{r\psi}$.
Once we keep these components then the metric \eqref{iibmetfinal} {\it cannot} be the global description. The global
description will have to be a non-geometric manifold. In the present
paper we will not discuss the non-geometric aspect anymore, and details on this will be presented in our upcoming
paper \cite{toappear}. We end this section by noting that
the duality cycle that we
advocated here (and also in \cite{gtpapers} earlier) does
lead to the correct gauge/gravity dualities for the confining theories.

\section{Discussions and conclusion}

In this paper we have made some progress along two related directions. The first one is to find a supersymmetric
configuration of wrapped D5-branes on a two-cycle of a resolved conifold. We find that the {\it simplest} way to
achieve supersymmetry here is to allow for a non-K\"ahler metric with an $SU(3)$ structure on the resolved conifold.
It is possible that once we add extra fluxes and fundamental seven-branes we can also
get a K\"ahler metric on the resolved conifold that is supersymmetric in the presence of the wrapped D5-branes.

The second progress that we made here is related to the full cycle of geometric transition. It should now be clear
that in the limit where the gauge theories on the wrapped D5-branes
or the D6-branes were
strongly coupled, the geometric transition
effectively  boiled down to a simple series of mirror transformations \cite{syz} followed
in-between by a flop transition to an intermediate M-theory configuration with a $G_2$-structure. The fact that
such a complicated set of gauge/gravity dualities follow from simple sequences of T-dualities is rather 
remarkable\footnote{An interesting question to ask at this stage is whether the set of operations (mirror symmetry
and flop) could be described as a transformation in the space of $SU(3)$ structures, much like the one studied in 
\cite{minabeta}. The fact that this is indeed the case for our set-up can be inferred from the torsion class
analysis before and after geometric transition. In the sequel \cite{chen2} we have worked out the torsion 
classes both before and after GT in type IIA theory. These torsion classes could be related to each other by 
$SU(3)$ transformations, and therefore our set of operations can be described as transformations in the space of 
$SU(3)$ structure.}.

In the global case the mirror transformations are more subtle than the local case.
For example the fluxes appearing in the mirror
pictures are very involved (we gave one example in {\bf Appendix 2}). They are not in general pure gauges, and so
we need to follow them through every step of the duality cycle because they would eventally influence the gravity
duals for the confining gauge theories on the wrapped D6- and D5-branes (at least in the far IR). In this
paper we have managed to work out these details, and showed that despite the complicated underlining flux-structures, the
gravity duals for the type IIA and the type IIB configurations in certain cases were
 given by non-K\"ahler deformations of the resolved and
the deformed conifold respectively. In more generic cases, that allow us to keep all components of the B-fields, the
gravitational descriptions become non-geometric.

There are many things that need to be done now. One of the issue is to understand the non-geometric aspects of the
mirror configurations. We have briefly alluded to this by arguing that there are B-field components (even at the
orientifold points) which could in-principle make the IIA, as well as IIB after geometric transition,
picture non-geometric (see also the recent discussion in
\cite{halmagyi}). Such a non-geometric configuration for example
in type IIA should lift to a non-geometric configuration in
M-theory. The question then is: how should we perform the flop operation now? Even if there exist a meaningful
way to perform the flop, how do we understand the SYZ \cite{syz} mirror map now? How does the non-geometric
behavior reflect in the gauge theory side? How should we compute the exact superpotential for the gauge theory in the
non-geometric case? Furthermore if we allow the B-fields components $\bar{b}_{\phi_1\phi_2}$ and
$\bar{b}_{\psi\phi_i}$ in \eqref{BIIA2}, how would the metric \eqref{iibmetfinal} change to reflect the non-geometric
aspect?

Clearly these are very interesting questions to resolve, and we have barely scratched the surface of the iceberg. In
our upcoming paper \cite{toappear} we plan to address some of these issues. Hopefully these questions
will have simple tractable
solutions. If not, then it will be even more interesting to reveal the underlying structure.

\vskip.1in

\noindent {\bf Note added:}
While we were writing this draft, \cite{pandozayas} appeared that has some overlap with sections 4.1 and
4.2 of this paper. See also the recent set of papers \cite{lilia} where some new properties of non-K\"ahler manifolds
have been discussed.

%\newpage

\vskip.1in

\begin{center}
{\bf Acknowledgements}
\end{center}

\noindent We would like to thank N.~Halmagyi, M.~Larfors, D.~L\"ust, and D.~Tsimpis for helpful
conversations. The works of FC, KD and PF were supported in parts by the NSERC grants.
The work of SK was supported by NSF grants DMS-02-44412 and
DMS-05-55678. The work of RT was supported by a PPARC grant.

\newpage
\appendix
\section{A more precise derivation of the initial metric}

We compute the metric using the formulas in Section~\ref{su3}.  The only thing left to
do is to choose coordinates.  We have already described complex coordinates $(z^2,z^3,z^4)$ in Section~\ref{rescon} so it only remains to convert these to real
coordinates.  We choose polar coordinates $(r_i,t_i)$ with $z_i=r_ie^{it_i}$,
and implement the computations in Section~\ref{su3}.

Before doing so, we implicitly describe
the coordinate change necessary to compare with the metric
(\ref{susybg}).  Noting that the $\bfC^*$ action on the homogeneous coordinates
of the resolved
conifold identifies $(z_1,z_2,z_3,z_4)$ with $(1,z_2/z_1,z_1z_3,z_1z_4)$, we
start with the coordinates $(U,Y,\lambda)$ of
\cite{pandoz}, which we see are related to our coordinates by
\bg\label{pztchange}
U=z_1z_3,\qquad Y=z_1z_4,\qquad \lambda=-\frac{z_2}{z_1}
\nd
due to a sign convention in \cite{pandoz}.  Using (\ref{resconcoords}) together
with (\ref{pztchange}),
we can express $(U,Y,\lambda)$ in terms of $(z_2,z_3,z_4)$, hence in terms of
the real coordinates $(r_2,r_3,r_4,t_2,t_3,t_4)$.  The desired
change of variables
comes from using (2.13) of \cite{pandoz}, which expresses $(U,Y,\lambda)$
instead in
terms of the desired six real variables $(r,\psi,\phi_1,\theta_1,\phi_2,\theta_2)$.

\bigskip
Ordering the coordinates as $(r_2,r_3,r_4,t_2,t_3,t_4)$, the components of the metric
are given as follows.

\begin{eqnarray}
g_{1,1}&=& \frac{1}{\left(2 r_3^2+2 r_4^2+u\right)^2}\left\{
-\frac{1}{\left(-r_2^2+r_3^2+
r_4^2+u\right)^2}\right.\nonumber\\
&&\times\left[\sqrt{-r_2^2+r_3^2+r_4^2+u} \left(a \left(2
r_3^2+r_4^2+u\right)-b^2 r_4^2\right)+r_2 r_3 r_4 \left(a+b^2\right)
\cos
(t_2-t_3+t_4)\right] \nonumber\\
&&\times\left[\sqrt{-r_2^2+r_3^2+r_4^2+u} \left(\left(2
r_3^2+u\right) \left(r_3^2+r_4^2+u\right)-2 r_2^2
\left(r_3^2-r_4^2\right)\right)\right.\nonumber\\
&&\left.-4 r_2 r_3 r_4
\left(r_2^2-r_3^2-r_4^2-u\right) \cos(t_2-t_3+t_4)\right]\nonumber\\
&&+\frac{r_2 \left(a+b^2\right)}{\left(-r_2^2+r_3^2+
r_4^2+u\right)^2} \left[r_2 r_3 \cos (t_2-t_3+t_4)-r_4
\sqrt{-r_2^2+r_3^2+r_4^2+u}\right] \nonumber\\
&&\times\left[r_2 r_4 \sqrt{-r_2^2+r_3^2+r_4^2+u} \left(-2 r_2^2+2
r_3^2+4 r_4^2+3 u\right)\right.\nonumber\\
&&\left.-2 r_3 \left(r_2^2-r_4^2\right)
\left(-r_2^2+r_3^2+r_4^2+u\right) \cos
(t_2-t_3+t_4)\right]\nonumber\\
&&+\frac{\left(a+b^2\right)}{\left(-r_2^2+r_3^2+
r_4^2+u\right)^{3/2}} \left[r_2 r_3-r_4 \sqrt{-r_2^2+r_3^2+r_4^2+u}
\cos
(t_2-t_3+t_4)\right] \nonumber\\
&&\times\left[r_2 r_3 \left(2 r_2^2-2 r_3^2-u\right)
\sqrt{u-r_2^2+r_3^2+r_4^2}\right.\nonumber\\
&&\left.+2 r_4 \left(r_2^4-3r_2^2r_3^2
+2r_4^2+(r_4^2+u)(u-2r_2^2+3r_3^2+r_4^2)\right) \cos
(t_2-t_3+t_4)\right]\nonumber\\
&&-\frac{2 r_2^2 r_3^2 \left(a+b^2\right) \left(r_2^2+r_4^2\right)
\sin^2(t_2-t_3+t_4)}{-r_2^2+r_3^2+r_4^2+u}\nonumber\\
&&\left.-2 r_4^2 \left(a+b^2\right) \left(-r_2^2+2 r_3^2+r_4^2+u\right) \sin^2(t_2-t_3+t_4)\right\}\nonumber\\
g_{1,2}&=& \frac{1}{\left(2 r_3^2+2 r_4^2+u\right)
\left(-r_2^2+r_3^2+r_4^2+u\right)}\left[r_2 r_3 \left(a
\left(r_2^2+r_4^2\right)+b^2 \left(-r_2^2+2
   r_3^2+r_4^2+u\right)\right)\right.\nonumber\\
&&\left.+r_4 \left(a-b^2\right) \sqrt{-r_2^2+r_3^2+r_4^2+u}
\left(-r_2^2+2 r_3^2+r_4^2+u\right) \cos
   (t_2-t_3+t_4)\right]\nonumber\\
g_{1,3}&=& \frac{1}{\left(2 r_3^2+2 r_4^2+u\right)^2}\left\{
\frac{1}{\left(-r_2^2+r_3^2+
r_4^2+u\right)^2}\right.\nonumber\\
&&\times\left[\sqrt{-r_2^2+r_3^2+r_4^2+u} \left(a \left(2
r_3^2+r_4^2+u\right)-b^2 r_4^2\right)+r_2 r_3 r_4 \left(a+b^2\right)
\cos
(t_2-t_3+t_4)\right]\nonumber\\
&&\times\left[r_2 r_4 \sqrt{-r_2^2+r_3^2+r_4^2+u} \left(-2 r_2^2+2
r_3^2+4 r_4^2+3 u\right)\right.\nonumber\\
&&\left.-2 r_3 \left(r_2^2-r_4^2\right)
\left(-r_2^2+r_3^2+r_4^2+u\right) \cos (t_2-t_3+t_4)\right]
\nonumber\\
&&-\frac{r_2 \left(a+b^2\right)}{\left(-r_2^2+r_3^2+r_
4^2+u\right)^2} \left[r_2 r_3 \cos (t_2-t_3+t_4)-r_4
\sqrt{-r_2^2+r_3^2+r_4^2+u}\right]\nonumber\\
&&\times \left[ 4 r_2
r_3 r_4 (r_2^2-r_3^2-r_4^2-u) \cos (t_2-t_3+t_4)+\sqrt{-r_2^2+r_3^2+r_4^2+u }  \right.\nonumber\\
&&\left. \times\left(2 r_2^4-r_2^2 \left(4 r_3^2+4 r_4^2+3
u\right)+2 r_3^4+r_3^2 \left(4 r_4^2+3 u\right)+\left(2
r_4^2+u\right)^2\right)\right]\nonumber\\
&&+\frac{\left(a+b^2\right)}{\left(-r_2^2+r_3^2+r_
4^2+u\right)^{3/2}} \left[r_2 r_3-r_4 \sqrt{-r_2^2+r_3^2+r_4^2+u}
\cos
(t_2-t_3+t_4)\right] \nonumber\\
&&\times\left[r_3 r_4 \left(u-2 r_2^2+2 r_3^2\right)
\sqrt{u-r_2^2+r_3^2+r_4^2}\right.\nonumber\\
&&\left.+2 r_2
\left(r_2^4-3r_2^2r_3^2+2r_4^2+(r_4^2+u)(u-2r_2^2+3r_3^2+r_4^2)\right)
\cos
(t_2-t_3+t_4)\right]\nonumber\\
&&-2 r_2 r_4 \left(a+b^2\right) \left(-r_2^2+2
r_3^2+r_4^2+u\right) \sin ^2(t_2-t_3+t_4)\nonumber\\
&&\left.+\frac{2 r_2 r_3^2 r_4 \left(a+b^2\right)
\left(r_2^2+r_4^2\right) \sin
^2(t_2-t_3+t_4)}{-r_2^2+r_3^2+r_4^2+u}\right\}\nonumber\\
g_{1,4}&=&\frac{r_2^2 r_3 r_4 \left(a-b^2\right) \sin
(t_2-t_3+t_4)}{\left(2 r_3^2+2
r_4^2+u\right) \sqrt{-r_2^2+r_3^2+r_4^2+u}}\nonumber\\
g_{1,5}&=&\frac{r_3 r_4 \left(a-b^2\right)
\sqrt{-r_2^2+r_3^2+r_4^2+u} \sin
(t_2-t_3+t_4)}{2 r_3^2+2 r_4^2+u}\nonumber\\
g_{1,6}&=& \frac{r_2^2 r_3 r_4 \left(a-b^2\right) \sin
(t_2-t_3+t_4)}{\left(2 r_3^2+2
r_4^2+u\right) \sqrt{-r_2^2+r_3^2+r_4^2+u}}\nonumber\\
g_{2,2}&=& \frac{\left(r_2^2-2 r_3^2-r_4^2-u\right) \left(a
\left(r_2^2+r_4^2\right)+b^2 \left(-r_2^2+2
r_3^2+r_4^2+u\right)\right)}{\left(2 r_3^2+2
r_4^2+u\right) \left(-r_2^2+r_3^2+r_4^2+u\right)}\nonumber\\
g_{2,3}&=&-\frac{1}{\left(2 r_3^2+2 r_4^2+u\right)^2}
\left\{\frac{r_4 \left(a+b^2\right)\left(-r_2^2+2
r_3^2+r_4^2+u\right) \cos (t_2-t_3+t_4)}{\left(-r_2^2+r_3^2+
r_4^2+u\right)^2}\right.\nonumber\\
&&\times\left[r_2 r_4 \sqrt{-r_2^2+r_3^2+r_4^2+u} \left(-2 r_2^2+2
r_3^2+4 r_4^2+3 u\right)\right. \nonumber\\
&&\left.-2 r_3 \left(r_2^2-r_4^2\right)
\left(-r_2^2+r_3^2+r_4^2+u\right) \cos
(t_2-t_3+t_4)\right]\nonumber\\
&&+\frac{r_2 \left(a+b^2\right)\left(r_2^2-2 r_3^2-r_4^2-u\right)
\cos (t_2-t_3+t_4)}{\left(-r_2^2+r_3^2+r_ 4^2+u\right)^2}
\nonumber\\
&&\times\left[4 r_2 r_3 r_4 \left(r_2^2-r_3^2-r_4^2-u\right) \cos
(t_2-t_3+t_4)+\sqrt{-r_2^2+r_3^2+r_4^2+u }\right.\nonumber\\
&&\left. \left(2 r_2^4-r_2^2 \left(4 r_3^2+4 r_4^2+3 u\right)+2
r_3^4+r_3^2 \left(4 r_4^2+3 u\right)+\left(2
r_4^2+u\right)^2\right)\right]\nonumber\\
&&+\frac{\left(b^2 \left(-r_2^2+2 r_3^2+r_4^2+u\right)-a
\left(r_2^2+r_4^2\right)\right)}{\left(-r_2^2+r_3^2+
r_4^2+u\right)^{3/2}} \left[r_3 r_4 \left(-2 r_2^2+2 r_3^2+u\right)
\right.\nonumber\\
&&\times\sqrt{-r_2^2+r_3^2+r_4^2+u}+2 r_2 \left(r_2^4-r_2^2
\left(3 r_3^2+2 \left(r_4^2+u\right)\right)+2 r_3^4\right.\nonumber\\
&&\left.\left.+3 r_3^2
\left(r_4^2+u\right)+\left(r_4^2+u\right)^2\right) \cos
(t_2-t_3+t_4)\right]\nonumber\\
&&+\frac{2 r_3 r_4^3 \left(a+b^2\right) \left(-r_2^2+2
r_3^2+r_4^2+u\right) \sin
^2(t_2-t_3+t_4)}{-r_2^2+r_3^2+r_4^2+u}\nonumber\\
&&\left.-\frac{2 r_2^2 r_3 r_4 \left(a+b^2\right) \left(r_2^2-2
r_3^2-r_4^2-u\right) \sin
^2(t_2-t_3+t_4)}{-r_2^2+r_3^2+r_4^2+u}\right\}\nonumber\\
g_{2,4}&=& \frac{r_2 r_4 \left(a-b^2\right) \left(r_2^2-2
r_3^2-r_4^2-u\right) \sin (t_2-t_3+t_4)}{\left(2 r_3^2+2
r_4^2+u\right)
\sqrt{-r_2^2+r_3^2+r_4^2+u}}\nonumber\\
g_{2,5}&=& 0\nonumber\\
g_{2,6}&=& \frac{r_2 r_4 \left(a-b^2\right) \left(r_2^2-2
r_3^2-r_4^2-u\right) \sin (t_2-t_3+t_4)}{\left(2 r_3^2+2
r_4^2+u\right)
\sqrt{-r_2^2+r_3^2+r_4^2+u}}\nonumber\\
g_{3,3}&=& -\frac{1}{\left(2 r_3^2+2 r_4^2+u\right)^2}
\left\{\frac{r_4 \left(a+b^2\right)}{\left(-r_2^2+r_3^2+
r_4^2+u\right)^2} \right.\nonumber\\
&&\times\left[r_2 \sqrt{-r_2^2+r_3^2+r_4^2+u}+r_3 r_4 \cos
(t_2-t_3+t_4)\right] \nonumber\\
&&\times\left[r_2 r_4 \sqrt{-r_2^2+r_3^2+r_4^2+u} \left(-2 r_2^2+2
r_3^2+4 r_4^2+3 u\right)\right.\nonumber\\
&&\left.-2 r_3 \left(r_2^2-r_4^2\right)
\left(-r_2^2+r_3^2+r_4^2+u\right)
\cos(t_2-t_3+t_4)\right]\nonumber\\
&&+\frac{1}{\left(-r_2^2+r_3^2+ r_4^2+u\right)^2}\left[4 r_2 r_3 r_4
\left(r_2^2-r_3^2-r_4^2-u\right) \cos
(t_2-t_3+t_4)\right.\nonumber\\
&&+\sqrt{-r_2^2+r_3^2+r_4^2+u } \left(2 r_2^4-r_2^2 \left(4 r_3^2+4
r_4^2+3 u\right)+2 r_3^4+r_3^2 \left(4 r_4^2+3
u\right) \right.\nonumber\\
&&\left.\left.+\left(2
r_4^2+u\right)^2\right)\right]\times\left[\sqrt{-r_2^2+r_3^2+r_4^2+u}
\left(a \left(-r_2^2+2 r_3^2+2 r_4^2+u\right)-b^2
r_2^2\right)\right.\nonumber\\
&&\left.-r_2 r_3 r_4 \left(a+b^2\right) \cos
(t_2-t_3+t_4)\right]\nonumber\\
&&+\frac{\left(a+b^2\right)}{\left(-r_2^2+r_3^2+
r_4^2+u\right)^{3/2}} \left(r_2 \sqrt{-r_2^2+r_3^2+r_4^2+u} \cos
(t_2-t_3+t_4)+r_3 r_4\right)\nonumber\\
&& \times\left(r_3 r_4 \left(u-2 r_2^2+2 r_3^2\right)
\sqrt{u-r_2^2+r_3^2+r_4^2}\right.\nonumber\\
&&\left.+2 r_2
\left(r_2^4-3r_2^2r_3^2+2r_4^2+(r_4^2+u)(u-2r_2^2+3r_3^2+r_4^2)\right)
\cos
(t_2-t_3+t_4)\right)\nonumber\\
&&-2 r_2^2 \left(a+b^2\right) \left(r_2^2-2
r_3^2-r_4^2-u\right) \sin ^2(t_2-t_3+t_4)\nonumber\\
&&\left.+\frac{2 r_3^2 r_4^2 \left(a+b^2\right)
\left(r_2^2+r_4^2\right) \sin
^2(t_2-t_3+t_4)}{-r_2^2+r_3^2+r_4^2+u}\right\}\nonumber\\
g_{3,4}&=& \frac{r_2 r_3 r_4^2 \left(b^2-a\right) \sin
(t_2-t_3+t_4)}{\left(2 r_3^2+2
r_4^2+u\right) \sqrt{-r_2^2+r_3^2+r_4^2+u}}\nonumber\\
g_{(3,5}&=& \frac{r_2 r_3 \left(a-b^2\right)
\sqrt{-r_2^2+r_3^2+r_4^2+u} \sin
(t_2-t_3+t_4)}{2 r_3^2+2 r_4^2+u}\nonumber\\
g_{3,6}&=& \frac{r_2 r_3 r_4^2 \left(b^2-a\right) \sin
(t_2-t_3+t_4)}{\left(2 r_3^2+2
r_4^2+u\right) \sqrt{-r_2^2+r_3^2+r_4^2+u}}\nonumber\\
g_{4,4}&=& -\frac{r_2^2 \left(a \left(-r_2^2+2
r_3^2+r_4^2+u\right)+b^2 r_4^2\right)}{2 r_3^2+2
r_4^2+u}\nonumber\\
g_{4,5}&=& -\frac{r_2 r_3 \left(a r_2 r_3
\sqrt{-r_2^2+r_3^2+r_4^2+u}-r_4 \left(a-b^2\right)
\left(-r_2^2+r_3^2+r_4^2+u\right) \cos (t_2-t_3+t_4)\right)}{\left(2
r_3^2+2
r_4^2+u\right) \sqrt{-r_2^2+r_3^2+r_4^2+u}}\nonumber\\
g_{4,6}&=& -\frac{b^2 r_2^2 r_4^2}{2 r_3^2+2
r_4^2+u}\nonumber\\
g_{5,5}&=& -\frac{r_3^2 \left(a \left(r_2^2+r_4^2\right)+b^2
\left(-r_2^2+r_3^2+r_4^2+u\right)\right)}{2 r_3^2+2
r_4^2+u}\nonumber\\
g_{5,6}&=& \frac{r_3 r_4 \left(a r_3 r_4
\sqrt{-r_2^2+r_3^2+r_4^2+u}-r_2 \left(a-b^2\right)
\left(r_2^2-r_3^2-r_4^2-u\right) \cos (t_2-t_3+t_4)\right)}{\left(2
r_3^2+2
r_4^2+u\right) \sqrt{-r_2^2+r_3^2+r_4^2+u}}\nonumber\\
g_{6,6}&=&-\frac{r_4^2 \left(a \left(-r_2^2+2
r_3^2+r_4^2+u\right)+b^2 r_2^2\right)}{2 r_3^2+2 r_4^2+u}
\end{eqnarray}

\newpage
\section{Components of G-fluxes in type IIA mirror configuration}

The twelve components of the four-form G-flux that we listed in \eqref{twelve} can be expressed in terms of the type
IIB metric and flux-components in the following way:


\newpage
\section{Metric components in type IIB after geometric transition}

The components of the final type IIB metric \eqref{iibmetfinal} can be expressed using the IIA components
$\bar{j}_{\mu\nu}$ in the following way:
\begin{eqnarray}
g_{\psi\psi}&=&1/\Big(\bar{j}_{\psi\psi}+\bar{\alpha}\Big(
\bar{j}_{\phi_1\phi_1}(\bar{b}_{\psi\phi_2}^2-\bar{j}_{\psi\phi_2}^2)+\bar{j}_{\phi_2\phi_2}(\bar{b}_
{\psi\phi_1}^2-\bar{j}_{\psi\phi_1}^2)\nonumber\\
&&+2\bar{j}_{\phi_1\phi_2}(\bar{j}_{\psi\phi_1}\bar{j}_{\psi\phi_2}
-\bar{b}_{\psi\phi_1}\bar{b}_{\psi\phi_2})+2\bar{b}_{\phi_1\phi_2}(\bar{b}_{\psi\phi_1}\bar{j}_{\psi\phi_2}
-\bar{j}_{\psi\phi_1}\bar{b}_{\psi\phi_2})\Big)\Big)\nonumber\\
g_{\psi\phi_1}&=&\bar{\alpha}(\bar{j}_{\phi_1\phi_2}\bar{j}_{\psi\phi_2}-\bar{j}_{\phi_2\phi_2}\bar{j}_{\psi\phi_1}
-\bar{b}_{\phi_1\phi_2}\bar{b}_{\psi\phi_2})/2\nonumber\\
g_{\psi\phi_2}&=&\bar{\alpha}(\bar{j}_{\phi_1\phi_2}\bar{j}_{\psi\phi_1}-\bar{j}_{\phi_1\phi_1}\bar{j}_{\psi\phi_2}
+\bar{b}_{\phi_1\phi_2}\bar{b}_{\psi\phi_1})/2\nonumber\\
g_{\phi_1\phi_1}&=&\left[4(\bar{j}_{\phi_2\phi_2}\bar{j}_{\psi\psi}
-\bar{j}_{\psi\phi_2}^2-\bar{b}_{\psi\phi_2}^2)+\bar{\alpha}\Big(\bar{b}_{\phi_1\phi_2}\bar{b}_{\psi\phi_2}
(\bar{j}_{\phi_1\phi_2}\bar{j}_{\psi\phi_2}-\bar{b}_{\phi_1\phi_2}\bar{b}_{\psi\phi_2})\right.\nonumber\\
&&\left.-\bar{j}_{\psi\phi_1}
\bar{j}_{\phi_2\phi_2}(\bar{j}_{\psi\phi_1}
\bar{j}_{\phi_2\phi_2}+\bar{b}_{\phi_1\phi_2}\bar{b}_{\psi\phi_2})+\bar{j}_{\phi_1\phi_2}\bar{j}_{\psi\phi_2}
(\bar{j}_{\psi\phi_1}\bar{j}_{\phi_2\phi_2}-\bar{j}_{\phi_1\phi_2}\bar{j}_{\psi\phi_2})\Big)\right]\nonumber\\
&&/4\Big(\bar{\alpha}^{-1}\bar{j}_{\psi\psi}+\bar{j}_{\phi_1\phi_1}(\bar{b}_{\psi\phi_2}^2-\bar{j}_{\psi\phi_2}^2)+\bar{j}_{\phi_2\phi_2}(\bar{b}_
{\psi\phi_1}^2-\bar{j}_{\psi\phi_1}^2)\nonumber\\
&&+2\bar{j}_{\phi_1\phi_2}(\bar{j}_{\psi\phi_1}\bar{j}_{\psi\phi_2}
-\bar{b}_{\psi\phi_1}\bar{b}_{\psi\phi_2})+2\bar{b}_{\phi_1\phi_2}(\bar{b}_{\psi\phi_1}\bar{j}_{\psi\phi_2}
-\bar{j}_{\psi\phi_1}\bar{b}_{\psi\phi_2})\Big)\nonumber\\
g_{\phi_2\phi_2}&=&\left[4(\bar{j}_{\phi_1\phi_1}\bar{j}_{\psi\psi}
-\bar{j}_{\psi\phi_1}^2-\bar{b}_{\psi\phi_1}^2)+\bar{\alpha}\Big(-\bar{b}_{\phi_1\phi_2}\bar{b}_{\psi\phi_1}
(\bar{j}_{\phi_1\phi_2}\bar{j}_{\psi\phi_1}+\bar{b}_{\phi_1\phi_2}\bar{b}_{\psi\phi_1})\right.\nonumber\\
&&\left.-\bar{j}_{\psi\phi_2}
\bar{j}_{\phi_1\phi_1}(\bar{j}_{\psi\phi_2}
\bar{j}_{\phi_1\phi_1}-\bar{b}_{\phi_1\phi_2}\bar{b}_{\psi\phi_1})+\bar{j}_{\phi_1\phi_2}\bar{j}_{\psi\phi_1}
(\bar{j}_{\psi\phi_1}\bar{j}_{\phi_1\phi_2}-\bar{j}_{\phi_1\phi_1}\bar{j}_{\psi\phi_2})\Big)\right]\nonumber\\
&&/4\Big(\bar{\alpha}^{-1}\bar{j}_{\psi\psi}+\bar{j}_{\phi_1\phi_1}(\bar{b}_{\psi\phi_2}^2-\bar{j}_{\psi\phi_2}^2)+\bar{j}_{\phi_2\phi_2}(\bar{b}_
{\psi\phi_1}^2-\bar{j}_{\psi\phi_1}^2)\nonumber\\
&&+2\bar{j}_{\phi_1\phi_2}(\bar{j}_{\psi\phi_1}\bar{j}_{\psi\phi_2}
-\bar{b}_{\psi\phi_1}\bar{b}_{\psi\phi_2})+2\bar{b}_{\phi_1\phi_2}(\bar{b}_{\psi\phi_1}\bar{j}_{\psi\phi_2}
-\bar{j}_{\psi\phi_1}\bar{b}_{\psi\phi_2})\Big)\nonumber\\
g_{\phi_1\phi_2}&=&\left[2\bar{\alpha}(\bar{b}_{\psi\phi_1}\bar{b}_{\psi\phi_2}
+\bar{j}_{\psi\psi}\bar{j}_{\phi_1\phi_2}-\bar{j}_{\psi\phi_1}\bar{j}_{\psi\phi_2})
+\bar{j}_{\psi\phi_2}\bar{j}_{\phi_1\phi_1}(\bar{j}_{\psi\phi_1}\bar{j}_{\phi_2\phi_2}
-\bar{j}_{\psi\phi_2}\bar{j}_{\phi_1\phi_2})\right.\nonumber\\
&&+\bar{j}_{\psi\phi_1}\bar{j}_{\phi_1\phi_2}(\bar{j}_{\psi\phi_2}\bar{j}_{\phi_1\phi_2}
-\bar{j}_{\psi\phi_1}\bar{j}_{\phi_2\phi_2})+\bar{b}_{\psi\phi_1}\bar{b}_{\phi_1\phi_2}(\bar{j}_{\phi_1\phi_
2}\bar{j}_{\psi\phi_2}-\bar{j}_{\psi\phi_1}\bar{j}_{\phi_2\phi_2}-\bar{b}_{\psi\phi_2}\bar{b}_{\phi_1\phi_2})\nonumber\\
&&\left.+\bar{b}_{\psi\phi_2}\bar{b}_{\phi_1\phi_2}(\bar{j}_{\phi_1\phi_
1}\bar{j}_{\psi\phi_2}-\bar{j}_{\psi\phi_1}\bar{j}_{\phi_1\phi_2})\right]/2\Big(\bar{\alpha}^{-1}\bar{j}_{\psi\psi}+\bar{j}_{\phi_1\phi_1}(\bar{b}_{\psi\phi_2}^2-\bar{j}_{\p
si\phi_2}^2)\nonumber\\
&&+\bar{j}_{\phi_2\phi_2}(\bar{b}_
{\psi\phi_1}^2-\bar{j}_{\psi\phi_1}^2)+2\bar{j}_{\phi_1\phi_2}(\bar{j}_{\psi\phi_1}\bar{j}_{\psi\phi_2}
-\bar{b}_{\psi\phi_1}\bar{b}_{\psi\phi_2})+2\bar{b}_{\phi_1\phi_2}(\bar{b}_{\psi\phi_1}\bar{j}_{\psi\phi_2}
-\bar{j}_{\psi\phi_1}\bar{b}_{\psi\phi_2})\Big)\nonumber\\
g_{\phi_1\theta_1}&=&\bar{\alpha}\Big(\bar{j}_{\psi\phi_2}(\bar{j}_{\psi\phi_1}\bar{b}_{\theta_1\phi_2}
+\bar{b}_{\psi\phi_1}\bar{j}_{\theta_1\phi_2}-\bar{j}_{\psi\theta_1}\bar{b}_{\phi_1\phi_2}-\bar{b}_
{\psi\theta_1}\bar{j}_{\phi_1\phi_2}-\bar{j}_{\psi\phi_2}\bar{b}_{\theta_1\phi_1})\nonumber\\
&&-\bar{b}_{\psi\phi_2}
(\bar{j}_{\psi\phi_1}\bar{j}_{\theta_1\phi_2}
+\bar{b}_{\psi\phi_1}\bar{b}_{\theta_1\phi_2}-\bar{j}_{\psi\theta_1}\bar{j}_{\phi_1\phi_2}-\bar{b}_
{\psi\theta_1}\bar{b}_{\phi_1\phi_2}-\bar{b}_{\psi\phi_2}\bar{b}_{\theta_1\phi_1})\nonumber\\
&&+\bar{j}{\phi_2\phi_2}
(\bar{b}_{\psi\theta_1}\bar{j}_{\psi\phi_1}-\bar{j}_{\psi\theta_1}\bar{b}_{\psi\phi_1}+\bar{b}_{\theta_1
\phi_1}\bar{j}_{\psi\psi})
+\bar{j}_{\psi\psi}(\bar{b}_{\phi_1\phi_2}\bar{j}_{\theta_1\phi_2}
-\bar{j}_{\phi_1\phi_2}\bar{b}_{\theta_1\phi_2})\Big)/2\nonumber\\
g_{\phi_2\theta_2}&=&\bar{\alpha}\Big(\bar{j}_{\psi\phi_1}(\bar{j}_{\psi\phi_1}\bar{b}_{\theta_2\phi_1}
+\bar{b}_{\psi\phi_2}\bar{j}_{\theta_2\phi_1}-\bar{j}_{\psi\theta_2}\bar{b}_{\phi_2\phi_1}-\bar{b}_
{\psi\theta_2}\bar{j}_{\phi_2\phi_1}-\bar{j}_{\psi\phi_1}\bar{b}_{\theta_2\phi_2})\nonumber\\
&&-\bar{b}_{\psi\phi_1}
(\bar{j}_{\psi\phi_2}\bar{j}_{\theta_2\phi_1}
+\bar{b}_{\psi\phi_2}\bar{b}_{\theta_2\phi_1}-\bar{j}_{\psi\theta_2}\bar{j}_{\phi_2\phi_1}-\bar{b}_
{\psi\theta_2}\bar{b}_{\phi_2\phi_1}-\bar{b}_{\psi\phi_1}\bar{b}_{\theta_2\phi_2})\nonumber\\
&&+\bar{j}{\phi_1\phi_1}
(\bar{b}_{\psi\theta_2}\bar{j}_{\psi\phi_2}-\bar{j}_{\psi\theta_2}\bar{b}_{\psi\phi_2}+\bar{b}_{\theta_2
\phi_2}\bar{j}_{\psi\psi})
+\bar{j}_{\psi\psi}(\bar{b}_{\phi_2\phi_1}\bar{j}_{\theta_2\phi_1}
-\bar{j}_{\phi_2\phi_1}\bar{b}_{\theta_2\phi_1})\Big)/2\nonumber\\
g_{\theta_1\theta_2}&=&\hat{j}_{\theta_1\theta_1}-\frac{\hat{j}_{\theta_1\phi_1}}{2}\nonumber\\
g_{\theta_2\theta_2}&=&\hat{j}_{\theta_2\theta_2}-\frac{\hat{j}_{\theta_2\phi_2}}{2} \nonumber\\
g_{\theta_1\theta_2}&=&\hat{j}_{\theta_1\theta_2}-\frac{\hat{j}_{\phi_1\phi_2}\hat{j}_{\theta_1\phi_1}\hat{j}_{\theta_2\phi_2}}{4\hat{j}_{\phi_1\phi_1}\hat{j}_{\phi_2\phi_2}} \nonumber\\
g_{rr}&=&\hat{j}_{rr}-g_{\phi_1\phi_1}g_{r\phi_1}^2-g_{\phi_1\phi_2}g_{r\phi_2}^2-\frac{\hat{j}_{r\psi}}{2\hat{j}_{\psi\psi}}\nonumber\\
g_{r\phi_1}&=&\frac{\hat{j}_{r\phi_1}g_{\phi_2\phi_2}-\hat{j}_{r\phi_2}g_{\phi_1\phi_2}
-\hat{j}_{r\psi}g_{\phi_1\phi_2}g_{\psi\phi_1}+\hat{j}_{r\psi}g_{\psi\phi_2}g_{\phi_1\phi_2}}
{2(g_{\phi_1\phi_1}g_{\phi_2\phi_2}-g_{\phi_1\phi_2})}\nonumber\\
g_{r\phi_2}&=&\frac{\hat{j}_{r\phi_2}g_{\phi_1\phi_1}-\hat{j}_{r\phi_1}g_{\phi_1\phi_2}
-\hat{j}_{r\psi}g_{\phi_1\phi_2}g_{\psi\phi_2}+\hat{j}_{r\psi}g_{\psi\phi_1}g_{\phi_1\phi_2}}
{2(g_{\phi_1\phi_1}g_{\phi_2\phi_2}-g_{\phi_1\phi_2})}\nonumber\\
g_{r\psi}&=&\frac{\hat{j}_{r\psi}}{2\hat{j}_{\psi\psi}}-g_{\psi\phi_1}g_{r\phi_1}-g_{\psi\phi_2}g_{r\phi_2}
\end{eqnarray}
The $\hat{j}_{i,j}$ components by which we expressed some of the above $g_{\mu\nu}$ components can be expressed in the
following way:


%\section*{References}

\end{document}